\documentclass{emulateapj}
\usepackage{graphicx,amsfonts}
\newcommand{\vlsr}{$V_{\rm LSR}$}
\newcommand{\kms}{km~s$^{-1}$}
\newcommand{\HI}{\ion{H}{1}}
\newcommand{\HII}{\ion{H}{2}}

\newcommand{\co}{{\rm $^{12}$CO}}
\newcommand{\jone}{($J=1\rightarrow0$)}

\newcommand{\ha}{\mbox{H$\alpha$}}

\begin{document}

\title{Giant Molecular Clouds in M33 \\
I -- BIMA All-Disk Survey}

\author{G. Engargiola, R. L. Plambeck, E. Rosolowsky, and L. Blitz}
\affil{Radio Astronomy Lab, University of California, Berkeley, CA 94720}
\email{greg@astron.berkeley.edu}

\begin{abstract}

We present the first interferometric \co~\jone~map of the entire
\ha~disk of M33. The $13''$ diameter synthesized beam corresponds to a
linear resolution of 50~pc, sufficient to distinguish individual giant
molecular clouds (GMCs).  From these data we generated a catalog of
148 GMCs with an expectation that no more than 15 of the sources are
spurious.  The catalog is complete down to GMC masses of $1.5 \times
10^{5}~M_{\sun}$ and contains a total mass of $2.3 \times
10^7~M_{\sun}$.  Single dish observations of CO in selected fields
imply that our survey detects $\sim 50$\% of the CO flux, hence that
the total molecular mass of M33 is $4.5 \times 10^{7}~ M_{\sun}$,
approximately 2\% of the \HI\ mass.  The GMCs in our catalog are
confined largely to the central region ($R<4$ kpc). They show a
remarkable spatial and kinematic correlation with overdense
\HI~filaments; the geometry suggests that the formation of GMCs
follows that of the filaments.  The GMCs exhibit a mass spectrum
$dN/dM \propto M^{-2.6\pm 0.3}$, considerably steeper than that found
in the Milky Way and in the LMC.  Combined with the total mass, this
steep function implies that the GMCs in M33 form with a characteristic
mass of $\sim 7 \times 10^4 M_{\sun}$.  More than 2/3 of the GMCs have
associated \ion{H}{2}~ regions, implying that the GMCs have a short
quiescent period.  Our results suggest the rapid assembly of molecular
clouds from atomic gas, with prompt onset of massive star formation.

\end{abstract}

\bigskip

\section{Introduction}

This paper presents the results of an interferometric survey in
\co~\jone ~ of the entire \ha~disk of M33. Observations were made with
the Berkeley-Illinois-Maryland Association array.  The proximity and
low inclination of M33, an SA(s)cd spiral in the Local Group, permits
direct and detailed comparisons of molecular, atomic, and ionized
hydrogen gas on the scale of individual GMCs.  The angular resolution
of the BIMA survey is $13''$, which corresponds to a linear resolution
of 50 pc at the distance of M33 \citep[0.85 Mpc, ][]{lfm93}.  This
resolution is adequate to separate, but not to resolve, individual
giant molecular clouds (GMCs) if they are similar in size to those in
the Milky Way \citep{psp3}.  A follow-up paper by \citep[][Paper
II]{rpeb03} describes a study of rotation properties, virial masses,
and morphology of 36 GMCs from this survey which were reobserved at
BIMA with 20 pc linear resolution.  The BIMA array is well-suited for
this extensive mapping project for three reasons.  (1) The 6-meter
diameter antennas have a large ($100''$ FWHM) field of view at 115
GHz; this keeps the number of pointing centers to a manageable level.
(2) With 10 telescopes, the array affords good snapshot imaging
capability; visiting each field only a few times per night yields a
clean synthesized beam.  (3) The SIS receivers \citep{ep98} have
excellent sensitivity; double sideband receiver temperatures at 115
GHz are typically 40 K.


Of the numerous mechanisms proposed for molecular cloud formation
\citep{eg90}, it is uncertain which, if any, is responsible for most
GMCs in spiral galaxies. Validating any cloud formation model requires
an unbiased survey of molecular clouds in a galaxy.  Prior to the
present study of M33, there existed no complete catalog of giant
molecular clouds in an external spiral galaxy.  Even cloud {\it
catalogs} in the Milky Way are incomplete because velocity blending
along the line-of-sight can prevent unambiguous separation into
individual clouds \citep[e.g.][]{issa90}.  Only in the Large Magellenic
Cloud, a Local Group irregular, has CO been surveyed with sufficient
resolution to measure accurately the GMC mass spectrum \citep{fukui}.

A decade ago, \citet{ws90} undertook a pioneering survey of
extragalactic GMCs in the nuclear region of M33.  With the Owens
Valley interferometer they mapped \co~\jone~ in 19 fields with a $1'$
field-of-view and a $7''$ synthesized beam.  Their data analysis
yielded a catalog of 38 GMCs with $(0.2 \mbox{ --- } 4) \times 10^{5}\
M_{\sun}$ and a mass distribution index similar to that of the Milky
Way.  Although their observations covered a significant fraction of
the central 2 kpc, they positioned most fields to include areas of
high optical extinction detected in CO emission with the NRAO 12 m
telescope, possibly biasing the cloud sample.  This paper attempts to
address this bias by presenting a complete survey of GMCs in M33 over
most of the optical disk.

\bigskip
\section{Observations}  

\begin{figure}
\plotone{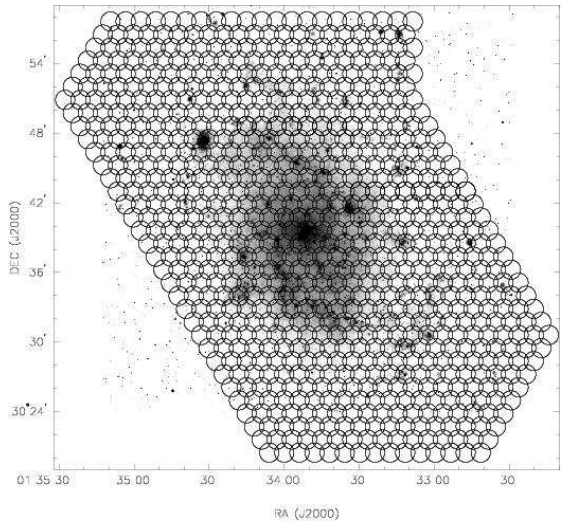}
\caption{Fields mapped with BIMA toward M33, overlaid on an H$\alpha$
image \citep{Cheng96} of the galaxy.  Circles show the $100''$ FWHM
primary beamwidth of the BIMA antennas at 115.27~GHz.}
\label{fig:pcenters}
\end{figure}

Observations were made with the BIMA array in its most compact (``D'')
configuration at various times between 1999 August and 2000 October.  As
shown in Figure~\ref{fig:pcenters}, M33 was mosaiced using a grid of 759
pointing centers, spaced by $78''$ in a hexagonal pattern.  Circles
indicate the FWHM primary beamwidth of the BIMA antennas, which is
$100''$ at 115.27~GHz.  The grid spacing was chosen to cover the galaxy
with as few pointing centers as possible while maintaining a point
source sensitivity that varies by no more than 10\% across the mosaic. 

The CO($1 \to 0$) line was observed in the upper sideband of the SIS
mixers.  Single sideband system temperatures, scaled to the top of the
atmosphere, were typically 300 to 500~K at 115.27~GHz.  The correlator
covered the LSR velocity range $-414$ to $+59$ \kms\ with 1.015~\kms\
channel spacing, on each of the 45 baselines. 

Each night we cycled through observations of about 50 grid points, using
an integration time of 25 seconds per point.  The slew and settle time
of the telescope between integrations was 5 seconds, so a complete loop
through 50 grid points required about 25 minutes.  Observations of the
calibrator, 0136$+$478, were interleaved with complete loops.  For a
typical 6-hour track, each pointing center was observed 10--12 times. 
With 45 baselines, this yields about 500 measurements of the source
visibility across the $(u,v)$ plane, resulting in a nearly round
synthesized beam with $\leq 10$\% sidelobes. 

A total of 45 observing tracks was obtained, with durations of a few
hours to 10 hours.  The total useful integration time accumulated on M33
was 132 hours, which corresponds to an average of 10.5 minutes per
pointing center. 

\subsection{Flux Scale}

For each observing track the amplitude gain of each antenna was scaled
by a constant factor to obtain the correct visibility amplitude on the
calibrator, 0136$+$478.  The flux density of this quasar was variable
over the course of the observations (2.3~Jy in 1999 September, 3.0~Jy
in 2000 May, 3.7~Jy in 2000 July, and 3.4~Jy in 2000 Oct).  
We monitored the calibrator flux density by making a brief
observation of the compact \ion{H}{2} region W3OH once per night.  The
flux density of W3OH was measured to be $3.8 \pm 0.2$ Jy by comparison
with primary flux calibrators Mars and Uranus.  All comparisons with
W3OH were made using the lower sideband continuum channel of the
receivers at 112~GHz, since measurements at 115 GHz are contaminated
by CO emission from the extended molecular cloud associated with W3OH.

The flux densities derived for 0136$+$478 were generally reproducible to
$\pm 0.2$~Jy from night to night.  Allowing for possible systematic
pointing errors and decorrelation by atmospheric phase fluctuations, we
estimate that the final amplitude scale is uncertain by approximately
$\pm 15$\%. 

\subsection{Mapping}

After calibrating the visibility data, we generated mosaiced spectral
line maps using the Miriad software
package\footnote{\url{http://www.atnf.csiro.au/computing/software/miriad}}. 
The maps were deconvolved with a CLEAN algorithm in order to remove
negative bowls around the strongest sources.  The synthesized beam
varies slightly across the mosaic because individual subgrids have
slightly different $(u,v)$ coverage.  To handle this, Miriad creates a
cube of synthesized beams, one for each pointing center, and
interpolates among them when deconvolving the map.  The maps were
restored with a $13''$ Gaussian. 

Most of our analysis was performed on maps with $2.03$~\kms\ velocity
resolution.  These were produced by Hanning smoothing the (1.015~\kms\
spacing) visibility spectra, then generating maps of alternate channels. 
The resulting channel maps are expected to be nearly
statistically independent. 

\begin{figure}
\plotone{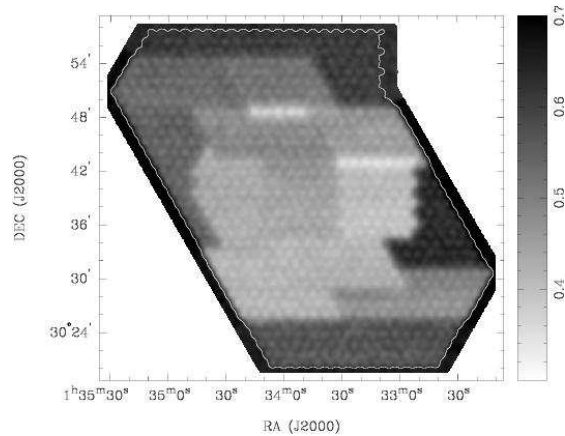}
\caption{RMS noise level, in Jy beam$^{-1}$, for the 2 \kms\
resolution CO data cube.  The white line near the edge of the mosaic
is the 99\% gain contour; outside this contour the data have not been
fully corrected for the primary beam response.  Within the 99\% gain
contour the RMS noise ranges from 0.34 to 0.65 Jy beam$^{-1}$; it is
$< 0.51$~Jy beam$^{-1}$ within the central half of the mosaic.}
\label{fig:rmsmap}
\end{figure}

To evaluate the noise level across the mosaic, we computed the RMS
fluctuations in the spectrum corresponding to each map pixel, omitting
any channels with flux densities $S > 4\sigma$.  From a spectrum with
222 statistically independent channels, the noise is determined to
approximately 5\% accuracy.  We then smoothed the map of the noise to
$30''$ resolution (Figure~\ref{fig:rmsmap}).  The white 99\% gain
contour denotes the effective outer boundary of the mosaic.  Outside
this contour intensities are not fully corrected for attenuation by
the primary beam.  For these 2 \kms\ resolution channel maps the RMS
noise level over the central half of the mosaic ranges from 0.34~Jy
beam$^{-1}$ to 0.51~Jy beam$^{-1}$; the average value is 0.44~Jy
beam$^{-1}$, or 0.24~K.  At the worst spots near the edge of the
mapped region (but inside the 99\% gain contour) the noise level is
sometimes as high as 0.65~Jy beam$^{-1}$.

\section{The GMC catalog}

\subsection{Identifying clouds}
\label{cloudfind}
\begin{figure}
\plotone{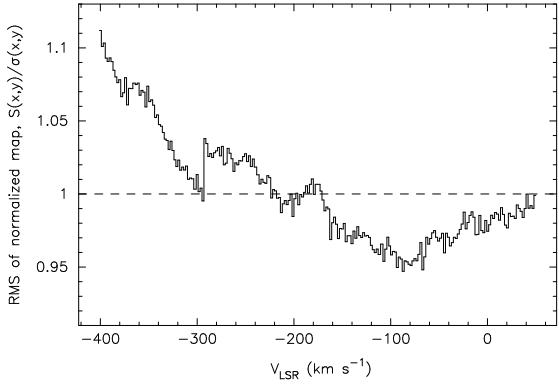}
\caption{RMS noise as a function of \vlsr, for the 2 \kms\ resolution
maps.  The noise level is frequency-dependent because of gain variations
across the instrumental passband.  In addition, the noise is higher at the
blueshifted end of the spectrum because of the rapid increase in atmospheric
opacity above 115~GHz.}
\label{fig:rmsvel}
\end{figure}

Because the sensitivity varies across the mosaic, we searched for
molecular clouds by clipping the maps at a constant signal-to-noise
ratio instead of a constant flux density threshold.  The intensity
data cube is converted to a signal-to-noise data cube by dividing
through by the noise at each position:
\begin{equation}
s(x,y,\nu) = \frac{I(x,y,\nu)}{\sigma(x,y) f(\nu)},
\end{equation}
where $\sigma(x,y)$ is the smoothed RMS noise map
(Figure~\ref{fig:rmsmap}) and $f(\nu)$ is a frequency-dependent noise
correction (Figure~\ref{fig:rmsvel}) which arises because of gain
variations of $5 - 10 \%$ across the instrumental passband.  Real CO
emission occupies only a few of the $\sim 20500$ independent resolution
elements in each channel map, thus having little effect on the
renormalization.  In the normalized, or signal-to-noise, data cube the
intensity is expressed as a multiple of the RMS noise at each pixel.
For example, $s(x,y,v) = -3$ represents a $3\sigma$ detection of a
negative peak.

\begin{figure}
\plotone{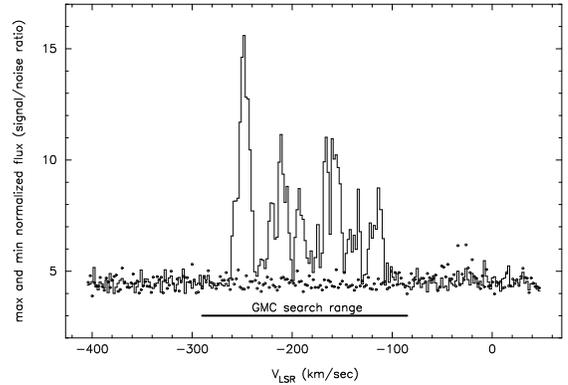}
\caption{Maximum and minimum normalized flux (signal-to-noise ratio) vs. 
\vlsr\ for the M33 CO data cube.  Maxima are indicated by the histogram,
minima by open circles.  The channel spacing is 2.03 \kms; the channels have
been boxcar averaged to obtain a velocity resolution of 4.06 \kms. 
Note that this is {\it not} the CO spectrum that one
would observe toward M33 with a sensitive single dish; in a large beam
the bright CO peaks would be greatly diluted by their small filling factors.}
\label{fig:4kmstat}
\end{figure}

\begin{figure}
\plotone{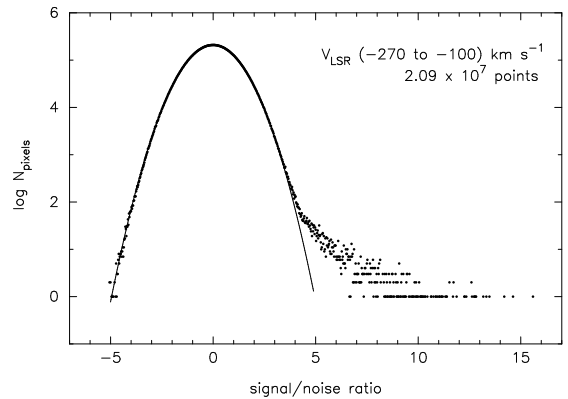}
\caption{Histogram of normalized flux values in the signal-to-noise
data cube, for 4 \kms\ channel maps, for the velocity range $-270 <
V_{\mathrm{LSR}} < -100$ \kms.  The bin size is 0.025, the total
number of points is 20916120, and the smooth curve is a Gaussian of
the form $y = 209048 \exp(-x^2/2)$.  The positive wing is due to CO
emission from molecular clouds.  Since the histogram plots $\sim 12$
dependent pixels for each statistically independent beam, the scatter in
the negative wing of the Gaussian is suppressed compared to if all the
points were independent.}
\label{fig:gstat2}
\end{figure}

The maximum and minimum normalized fluxes in each velocity channel are
plotted in Figure~\ref{fig:4kmstat}.  CO emission is obvious in the
range $-260 < V_{\mathrm{LSR}} < -105$ \kms.  A histogram of the
normalized fluxes in this velocity range (Figure~\ref{fig:gstat2}) is
well-fitted by a normal distribution, apart from the wing of high
signal-to-noise pixels attributable to CO detections.

Clearly, one could comfortably accept as a CO detection any pixel with
normalized flux $> 5 \sigma$.  Unfortunately, such a simple strategy
would miss much real molecular emission -- Figure~\ref{fig:gstat2}
shows a clear excess in pixels with flux $> 3.5 \sigma$.  Since the
velocity maps are nearly statistically independent, we can obtain a
higher detection efficiency by searching for neighboring channels with
unusually high normalized fluxes.  The lowest mass molecular cloud
which our survey is likely to detect is of order $2\times 10^{4}~
M_{\sun}$.  A typical Milky Way GMC with this mass has a velocity
dispersion of 1.8 \kms\ \citep{srby87}, which corresponds to a FWHM
linewidth of 4.2 \kms.  Therefore, we expect that CO emission from
most real clouds should appear in at least two adjacent 2 \kms\
velocity channels.  To identify molecular cloud candidates, we wrote
an IDL routine to locate pairs of adjacent channels with normalized
flux $> 2.7\sigma$.  Each pair, and all contiguous pixels with
normalized flux $> 2\sigma$, are assigned to a candidate molecular
cloud.  These pixels are then masked to avoid double counting, and a
search for another candidate cloud proceeds until all pairs of
adjacent $> 2.7\sigma$ channels have been masked.  The assignment of
pixels to candidate clouds in this manner does not depend on search
order but occasionally links nearby sources which could be regarded as
separate clouds.  Although we expect from our map statistics to find
real CO emission only in the velocity range $-280 < V_{\mathrm{LSR}} <
-85$ \kms, the search was conducted over the full 450 \kms\ range of
the data cube.  A total of 1600 candidate sources were identified in
this way.

To evaluate the likelihood that each candidate source is real, we
examine the most significant spectrum through the cloud -- either the
one with the highest summed significance or the one with the greatest
number of adjacent channels with flux $> 2\sigma$.  The calculation
described in Appendix~A evaluates the probability $P$ that this
spectrum is a noise fluctuation, {\it i.e.} that the detection is
spurious.  As discussed in Appendix~A, the likelihood that the cloud
is real is then given by $\mathrm{exp}(-NP) \sim 1 - NP$, where $N$ is
the number of statistically independent ways in which $n$ adjacent
channels can be selected from the data cube.  For the full data cube
$N \sim n_{beams} \cdot n_{chans} \sim 4.5\times 10^6$.  Note that
the likelihood computation is based on a single spectrum through each
cloud, not on an average over the source.  This keeps the calculation
simple, since adjacent spatial pixels oversample the synthesized beam
and hence are not statistically independent.

Originally, we intended to catalog all candidate sources with
$\lesssim 10\%$ likelihoods of being noise fluctuations.  To our
disappointment, however, the algorithm identified several sources
which were almost certainly spurious -- i.e., well outside the
plausible range for CO emission -- with false detection likelihoods of
$< 0.001$.  There are several possible explanations for this
inconsistency.  First, there may be outliers caused by a few bad
visibility measurements, perhaps from observations obtained in
especially poor weather or at low elevation.  Second, we may be
underestimating the number $N$ of statistically independent resolution
elements in the data cube.  Nyquist sampling the highest spatial
frequency visibility data (corresponding to antenna separations of 11.5
k$\lambda$) requires a $9''$ interval, which would imply there are
$\sim 50000$ independent resolution elements in the map.  We believe
the actual number is close to our original estimate of 20500 because
the long baseline data receive little weight in the final image.
Third, there may be errors in normalizing the maps by the rms noise.
Such errors are particularly likely for the frequency dependent
correction (Figure \ref{fig:rmsvel}) because there were slow changes in
the passband gain over the 14 month period in which observations were
made.

\begin{figure}
\plotone{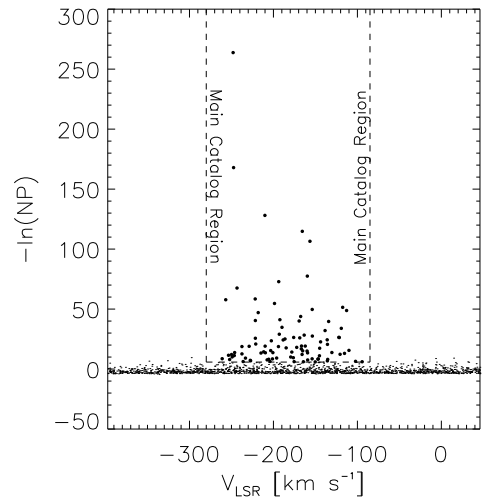} 
\caption{Likelihood measure $\ln(NP)$ as a function of
$V_{\mathrm{LSR}}$ for the 1600 cloud candidates.  Large values of
$-\ln(NP)$ represent the most significant clouds.  Only detections
within the velocity range $-280 \mbox{ km s}^{-1} < V_{\mathrm{LSR}} < -85
\mbox{ km s}^{-1}$ are included in the catalog.  The remainder of the
velocity range is used to estimate the number of false detections at a
given significance.  Clouds included in the main catalog are marked
with large dots.}
\label{fig:prob_vel}
\end{figure}

Because the likelihood calculation is imperfect, we use it only to
rank the {\it relative} likelihoods of the candidate sources.  For
convenience we define a `likelihood measure' as $-\ln(NP)$.
Figure~\ref{fig:prob_vel} plots this quantity vs. LSR velocity for the
1600 candidate sources.  Sources with the 60 highest likelihood
measures ($-\ln(NP) > 13.4$) fall in the velocity range $-260 <
V_{\mathrm{LSR}} < -105$ \kms and are apparent in multiple channel
maps, i.e.  they are unquestionably real.  As the likelihood threshold
is lowered, we begin to find a few sources that probably are spurious,
since they lie well outside this velocity range.  At $-\ln(NP) \sim 0$
the number of obviously spurious sources rises sharply, indicating the
noise floor of the data cube.

For our primary GMC catalog, we accept all cloud candidates in the
velocity interval $-280 < V_{\mathrm{LSR}} < -85$ \kms\ for which
$-\ln(NP) > 5.83$.  A total of 93 sources fit these criteria.  It is
straightforward to estimate the number of false detections in this
catalog simply by counting the number of sources above the likelihood
cutoff which lie outside the 96-channel search range.  There are 3
spurious detections in the 48 channels just below the search range,
and 3 spurious detections in the 48 channels just above the search
range.  Hence, we expect that there are of order 6 spurious detections
in this catalog.

\begin{figure}
\plotone{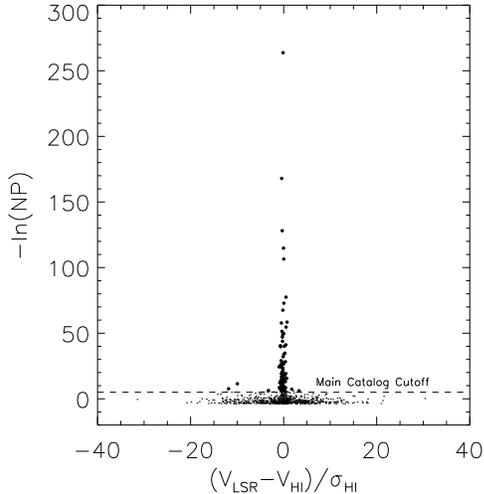}
\caption {Likelihood measure vs. the CO--\ion{H}{1} velocity
difference, normalized by the \ion{H}{1} linewidth.  The most
significant detections are within 3 velocity widths of the \ion{H}{1}
velocity.}
\label{fig:prob_widths}
\end{figure}

One can identify fainter sources by narrowing the search range to
cover only velocities that lie within the \ion{H}{1} emission line
profile in each direction.  Figure ~\ref{fig:prob_widths} shows the
cloud likelihoods from Figure~\ref{fig:prob_vel} plotted as a function
of the \HI--CO velocity difference.  \HI\ velocities were obtained
from the Westerbork map of \citet{deul87}, which has $12 \times 24''$
resolution.  For the 93 clouds in the primary catalog, the velocity
difference exceeds $\sigma_{\mathrm{HI}}$ for only 5 sources, where
$\sigma_{\mathrm{HI}}$ is the velocity dispersion of the \HI\ line.
The four sources with velocity differences $> 2\sigma_{\mathrm{HI}}$
may be false detections.

\begin{figure}
\plotone{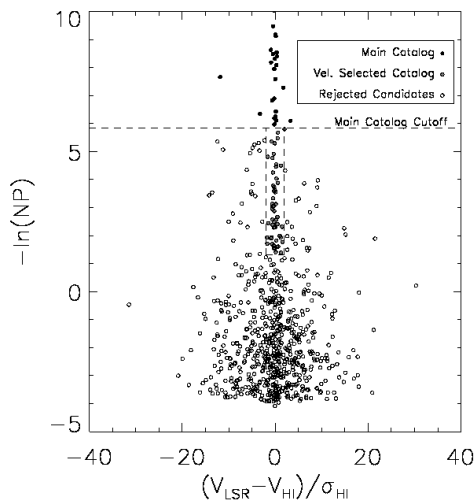} 
\caption {The 93 clouds in the main catalog are supplemented by 55
detections selected by their proximity to the \ion{H}{1} velocity.
There is a significant overdensity of clouds near the \ion{H}{1}
velocity, reflecting the presence of real clouds that are
indistinguishable from noise.  Because of the narrow velocity range,
we estimate that fewer than 9 of these clouds are false detections.  }
\label{fig:vsel}
\end{figure}

Narrowing the search range to $\pm 2\sigma_{\mathrm{HI}} \sim \pm 15$
\kms\ from the \HI\ velocity, we identified 55 additional clouds with
likelihood measure $-\ln(NP) > 1.4$.  Figure~\ref{fig:vsel} shows
these sources on a likelihood--velocity difference plot.  Once again,
by counting the number of (presumably) spurious sources outside the
velocity search range, we estimate that there are of order 9 false
detections in this catalog extension.

It is not useful to catalog individual clouds with likelihoods $-
\ln(NP) < 1.4$ because the number of spurious sources is quite large.
Nevertheless, in Figure~\ref{fig:vsel} there is an obvious overdensity
of GMC candidates with \vlsr\ within $2\sigma_{HI}$ of
$V_{\mathrm{HI}}$, and this may be used to gauge the number of cloud
candidates which are falsely rejected.  There are 145 candidate
sources with $2 < \mid V_{\mathrm{CO}}-
V_{\mathrm{HI}}\mid/\sigma_{\mathrm{HI}} < 4$ which are presumably
spurious.  One expects to find an equal number of spurious sources in
the range $\mid
V_{\mathrm{CO}}-V_{\mathrm{HI}}\mid/\sigma_{\mathrm{HI}} < 2$.  Since
there are actually 205 candidate sources in this range, we estimate
that roughly 60 clouds have been falsely rejected from our GMC catalog
in this significance range.  This result is not particularly sensitive
to how we select boundaries for the overdensity band in
Figure~\ref{fig:vsel}.

\subsection{Cloud Positions, Velocities, and Masses}

\begin{figure*}
\plotone{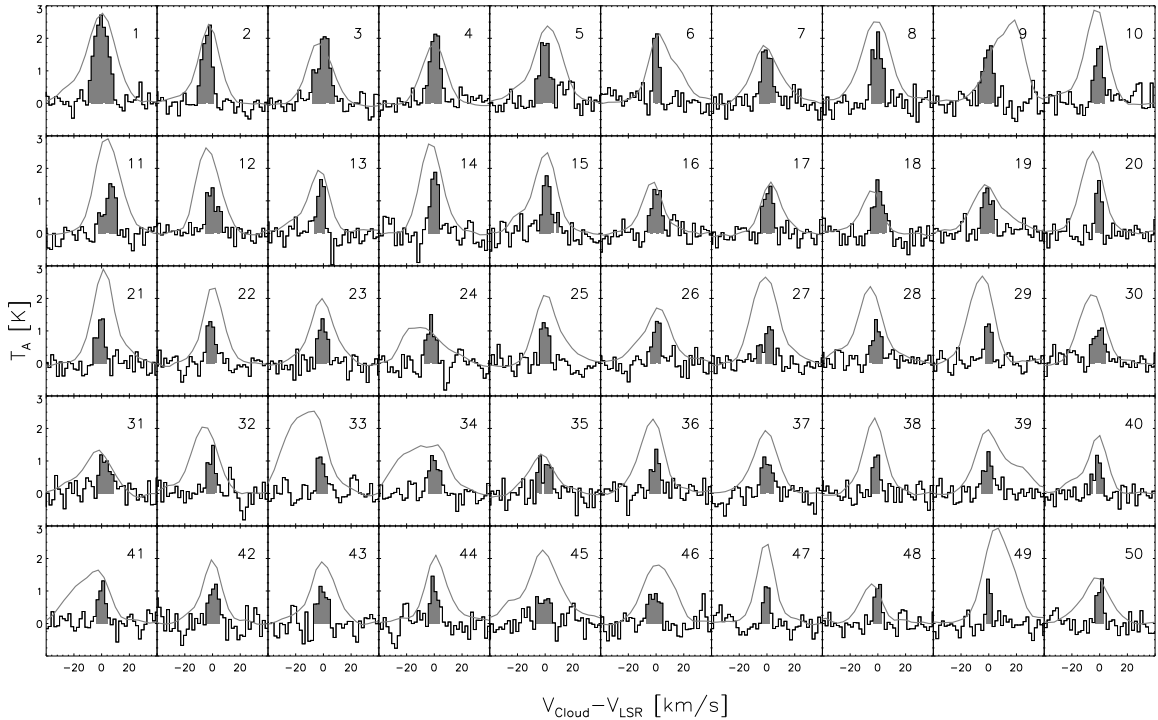}
\caption{Spectra for the first 50 clouds in the catalog.  The channels
included by the cloud identification algorithm are shaded and the grey
line represents the \ion{H}{1} line profile, scaled to have equal
amplitudes for the first cloud.}
\label{fig:cloudspectra}
\end{figure*}

Figure~\ref{fig:cloudspectra} shows the most significant CO spectrum
through each cloud, overplotted on the \HI\ spectrum in the same
direction.  Table \ref{cloudprop} lists the positions and properties of
the 148 clouds identified in our survey, ranked by likelihood.  Sources
1--93 were chosen without regard to the CO--\HI\ velocity difference,
and hence are suitable for statistical tests against the \HI\
distribution. 

\begin{table}
\begin{tabular}{c}
\hline\hline
\caption{Summary of cloud parameters for
HCRO observations.\label{cloudprop}}\\
\hline
Table omitted for brevity. The table is available at \\
\verb+http://astro.berkeley.edu/~eros/m33.table.txt+\\
\hline
\end{tabular}
\end{table}

The position and velocity of each cloud listed in Table \ref{cloudprop} are
intensity-weighted averages over all pixels which constitute the cloud
{\it i.e.,} $\bar{x}=(\sum T x)/(\sum T)$.  The velocity width is derived
from the intensity-weighted second moment of the velocity
distribution, 
\begin{equation}
\Delta v_{\rm FWHM} = \left[
8\,\mathrm{ln}2\,\frac{\sum T (v-\bar{v})^2}{\sum T} \right]^{0.5}.
\end{equation}
To estimate the cloud's mass, we convert the CO integrated intensity
for each pixel to an H$_2$ column density $N(\mbox{H}_2)= T(x,y,v)
\Delta v \cdot X$, where $\Delta v$ is the channel velocity width,
2.032 \kms, and where we assume a conversion factor $X = 2 \times
10^{20}~ \mbox{cm}^{-2}/(\mbox{K~km~s}^{-1})$
\citep{srby87,strong88,sm96,digel99}. The higher resolution
observations presented in Paper II imply that this conversion factor
is consistent with the virial masses of clouds out to radii of 4.5
kpc, despite variations in metallicity.  Each pixel corresponds to a
physical area $D^2 \Delta \alpha \Delta \delta$, where $D$ is the
distance to M33, 850 kpc \citep[][and references therein]{lfm93}, and
$\Delta \alpha$ and $\Delta \delta$ are the extent of the pixel in
angle, both of which are $4''$.  Thus, the total cloud mass is
\begin{equation} 
M = \sum_{\mathrm{pixels}} 1.36 \cdot 2 \cdot m_H \cdot X \cdot T_{\mathrm{
pixel}} \Delta v \cdot D^2 \Delta \alpha \Delta \delta.
\end{equation}
The factor 1.36 is the mass correction for helium with a number
fraction of 9\% relative to hydrogen nuclei. If expressed in terms of
flux density rather than brightness temperature, this is equivalent to
\begin{equation}
M = 7.5 \times 10^3 M_{\sun}/(\mbox{Jy km s}^{-1}).
\end{equation} 
This is 25\% lower than the conversion factor adopted by \citet{ws90},
mostly because of our choice of a lower $X$-factor.

\subsection{Completeness}
\label{2sigma}

To gauge the completeness of the survey, we added a regular lattice of
artificial test sources to the cube of channel maps and computed the
fraction of them recovered by our search algorithm.  Each test source
was a $13'' \times 13''$ Gaussian, corresponding to a GMC unresolved by
our synthesized beam, with a Gaussian velocity profile.  The velocity
width was kept constant at 6 \kms\ FWHM, independent of the flux density
(mass) of the test source.  Artificial sources were spaced by $100''$ in
right ascension and declination, over the entire mosaic; and by 56.3
\kms\ in velocity, from $-345$ to $-11$ \kms.  Sources close to the edge
of the mosaic were omitted if more than 5\% of their flux lay outside
the mosaic's 99\% gain contour.  In total, 2625 artificial sources were
added to the data cube.  Source recovery tests were done with masses
ranging from $3 \times 10^4~M_{\sun}$ to $10^6~M_{\sun}$. 

\begin{figure}
\plotone{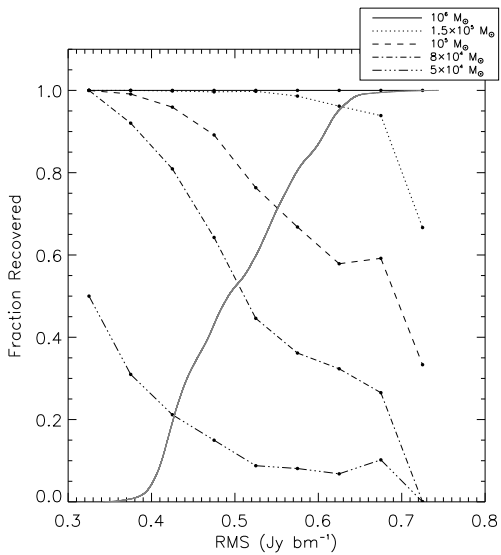}
\caption{Fraction of artificial clouds recovered as a function of RMS
noise, for 5 cloud masses.  The solid curve shows the cumulative
distribution of noise in our mosaic, within the 99\% gain contour.
Clouds with masses $> 1.5 \times 10^5~M_{\sun}$ will be detected
virtually anywhere in the map; most clouds with mass $> 1.0 \times
10^5~M_{\sun}$ will be detected within the inner half of the mosaic.}
\label{fig:num_recovered}
\end{figure}

Figure~\ref{fig:num_recovered} shows the fraction of the test sources
detected as a function of the local RMS noise level.  Over the full
mosaic, the survey is complete to $1.5 \times 10^{5}~M_{\sun}$.  Within
the central half of the mosaic, where the noise is typically $< 0.5$ Jy
beam$^{-1}$, we recover approximately 90\% of $1 \times 10^5~M_{\sun}$
clouds.  Outside this region, we recover 70 --- 80\% of such clouds.

\begin{figure}
\plotone{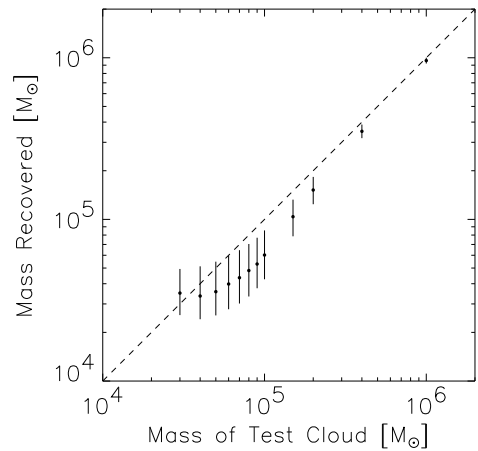}
\caption{Mass recovered as a function of test cloud mass.  Inverting
this relationship for masses above $5\times 10^4\ M_{\odot}$
determines the correction factor applied to the masses in Table
\ref{cloudprop}.}
\label{fig:frac_recovered}
\end{figure}

Figure~\ref{fig:frac_recovered} shows the recovered mass versus the
actual mass of the test sources.  Our algorithm underestimates cloud
masses because it integrates the CO luminosity only over pixels inside
the 2$\sigma$ brightness contour.  The underestimates are particularly
severe for low mass clouds, for which much of the emission is beneath
below the 2$\sigma$ noise level.  For clouds at the threshold of
detectability, $\sim 7 \times 10^4~M_{\sun}$, we recover about half
the mass, whereas for clouds more massive than $2 \times
10^5~M_{\sun}$, we recover more than 80\%.  The average correction
factors from Figure~\ref{fig:frac_recovered} were applied to our
catalog sources to obtain more accurate mass estimates; these are
given as $M_{\mathrm{CO}}^*$ in column (8) of Table 1.

\subsection{Extended sources}
\label{sens_budget}


The completeness estimates discussed above are applicable only to
sources that are unresolved by the $13''$ synthesized beam, i.e.  to
GMCs $\ll 50$ pc in diameter.  Because an aperture synthesis
telescope acts as a spatial filter, attenuating low spatial
frequencies, the sensitivity to larger sources is reduced.
Essentially, the synthesized beam consists of a narrow positive lobe
surrounded by a broad negative bowl.  If one convolves an extended
source brightness distribution with this beam, the negative
contribution from the bowl partially cancels the positive contribution
from the central lobe, reducing the observed flux density.  For the
BIMA D-array, the observed flux density for a $40''$ FWHM Gaussian
source is about half the true flux density.  Thus, our survey may not
detect a group of molecular clouds with a total mass of $1 \times
10^5~M_{\sun}$ spread out over a $40''$ (150 pc) region, though it is
quite likely to detect a single GMC with this mass.  For sources with
radii comparable to the beam size, their surface brightness will be
diminished relative to that of a point source by a factor of $\sim 2$.
Fortunately, sources of this size are also significantly more massive
($10^6~M_\odot$, Paper II) than our completeness limit and would be
readily detectable.

\begin{figure}
\plotone{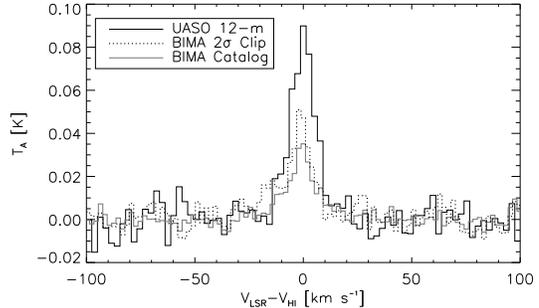}
\caption{A comparison of the flux selected in the D-array survey with
that observed by the UASO 12-m telescope.  A composite spectrum (solid
line) is shown for 15 single dish fields on a $7.5'$ cut along the
major axis of M33.  This is derived by shifting the spectrum for each
of the fields to a common velocity and averaging.  The integrated
intensity of the 12 m spectrum is 1.2~K~km~s$^{-1}$ corresponding to
$\sim$ 5.2 M$_\sun$ pc$^{-2}$.  For comparison, composite spectra are
shown for BIMA data within the 12~m fields, convolved to the $54''$
FWHM 12~m beam.  Two BIMA spectra were synthesized, one from the
catalog source pixels only (thick line), another from the data cube
clipped at 2$\sigma$ (dotted line).  The BIMA catalog spectrum
contains 43\% of the single dish flux, the clipped spectrum, 54\%.  }
\label{fig:censpec}
\end{figure}

Single dish observations of M33 with the University of Arizona Steward
Observatory 12-m telescope were used to estimate the total fraction of
the CO flux recovered by the BIMA survey.  The single dish flux
calibrations were repeatable within 6\%, and agreed with BIMA flux
measurements of planetary and spectral line calibrators within 10\%;
further observational details are given in Paper II.  Two sets of
single dish observations were made.  First, 18 positions with BIMA
catalog sources were observed.  For these fields the BIMA D-array maps
recovered approximately 60\% of the single dish CO flux.  CO was
detected in just 2 of the 18 {\it off-source} spectra obtained in
these position-switched observations; the 16 non-detections may be
used to set an upper limit of 0.3 M$_\sun$ pc$^{-2}$ for the surface
density of an extended ($R \leq 4$ kpc) uniform molecular disk.
Second, we observed 15 fields to map a $7.5'$ cut along the major axis
of M33.  In Figure~\ref{fig:censpec} we show a composite single dish
spectrum derived by shifting the spectrum for each of the 15 fields to
a common velocity and averaging.  The integrated flux of the of the 12
m spectrum is 1.2~K~km~s$^{-1}$ corresponding to $\sim$5.2 M$_\sun$
pc$^{-2}$.  This corresponds to $1.6\times 10^{7}~M_{\sun}$ of
molecular gas in the central $7'$ diameter of the galaxy.
\citet{ws89} measure $1.9\times 10^{7}~M_{\sun}$ in the same region
when converted to our $X$-factor, distance, and removing their
correction for source-beam coupling.

For comparison between the 12~m and the interferometer, we show
composite spectra made from BIMA data within the 12~m fields,
convolved to the $54''$ FWHM 12~m beam, sampled at the centers of the
12~m beams, and averaged together in a fashion identical to the 12~m
spectra.  Two BIMA spectra were synthesized, one from the catalog
source pixels only (thick line), another from the data cube clipped at
2$\sigma$ (dotted line).  The BIMA catalog spectrum contains 43\% of
the single dish flux, the clipped spectrum, 54\%.  In Paper II we
argue that most of the missing flux is attributable to complexes of
undetected low mass clouds associated with catalog sources.

We can estimate the total molecular mass of M33 from the major axis
cut.  The integrated single dish molecular mass in this $7.5'$ long
strip is $2.8 \times 10^6~M_\sun$.  The corresponding BIMA mass,
considering only cataloged clouds with masses greater than $1.5 \times
10^5~M_\sun$, the completeness limit for our survey, is $9.2 \times
10^5$~M$_\sun$, 33\% of the single dish mass.  In this computation we
have weighted the cloud masses by the appropriate 12-m beam
attenuation factors.  The total (corrected) mass in the catalog is
$2.3 \times 10^7~M_{\sun}$, of which 1.5$\times10^{7}~M_{\sun}$ is
from clouds more massive than the completeness limit.  If the GMC mass
spectrum in the central cut is similar to that for the galaxy as a
whole, then we infer that the total molecular mass is 3 times this
value, or 4.5$\times10^{7}~M_{\sun}$.

\subsection{Comparison with Wilson and Scoville 1990}

\begin{figure}
\plotone{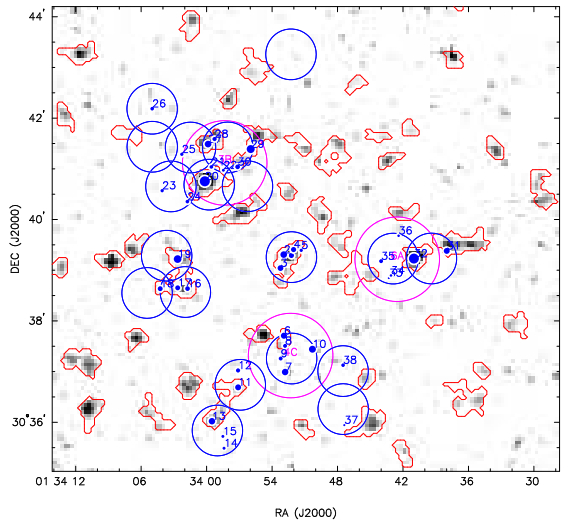}
\caption{Comparison of BIMA and WS90 molecular clouds in M33.  The map
covers the central $10' \times 10'$ region of M33.  The gray scale
shows a BIMA D-array image generated by masking out all emission $<
2\sigma$ in individual 2.03 \kms channel maps, then integrating over
the velocity interval $-240 <$ \vlsr$ < -128$ \kms.  Red contours show
the outer boundaries of BIMA clouds determined by our search
algorithm.  Open blue circles show the 19 fields mapped by WS90;
sources cataloged by WS90 are shown as filled blue circles, with area
proportional to mass.  Magenta circles indicate the 3 fields mapped in
the BIMA C-array, as described in the text.}
\label{fig:WS2}
\end{figure}

\citet[][hereafter WS90]{ws90} used the Owens Valley array to map 19
fields in the nucleus of M33 with $7''$ resolution, identifying 38
molecular clouds.  Figure~\ref{fig:WS2} is an overlay of the WS90
fields on the central $10' \times 10'$ section of the BIMA survey.
Red contours show the outer boundary of clouds identified in the BIMA
survey; solid circles mark locations of clouds cataloged by WS90.
Frequently, a BIMA cloud corresponds to two or more WS clouds -- for
example, the complex WS1-WS5 at the center of the galaxy is classified
as one cloud in our survey because these sources are connected by
low-level emission and lower BIMA resolution blends these clouds
together.  The BIMA survey detects all of the WS90 clouds with masses
$> 2 \times 10^5 M_{\odot}$.

\begin{figure*}
\epsscale{0.7}
\plotone{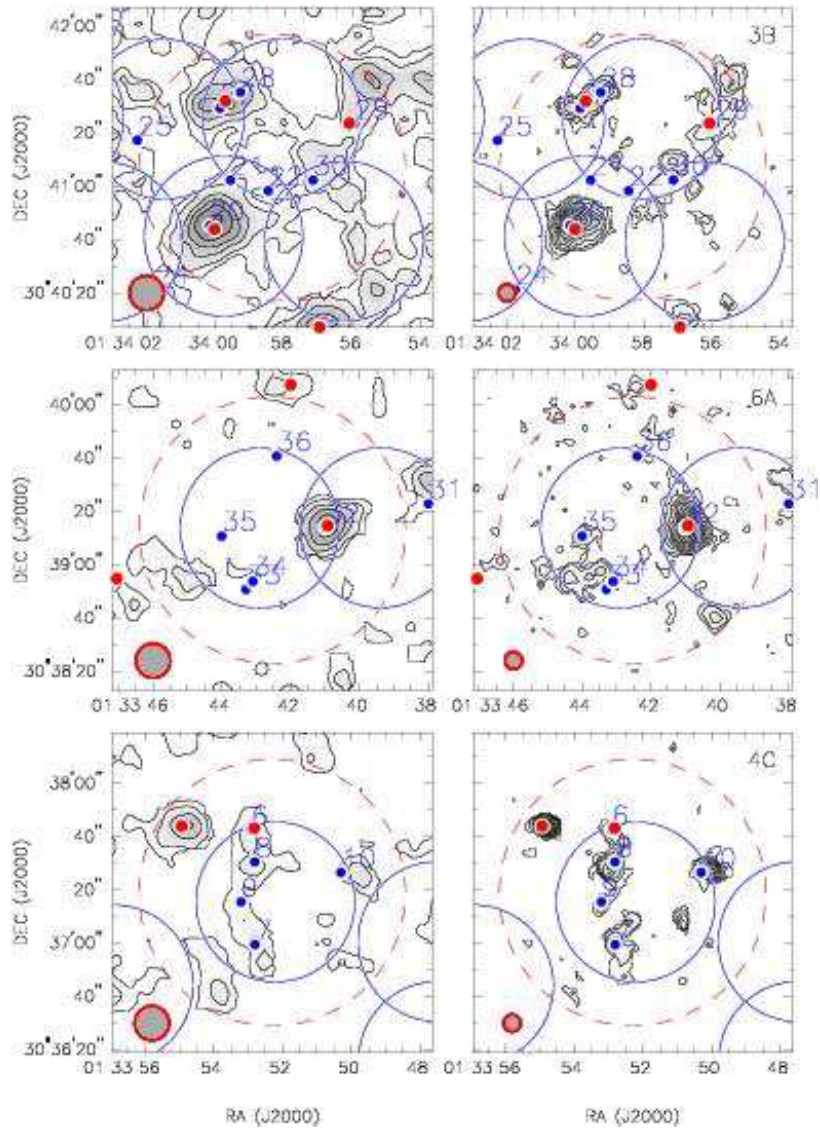}
\caption{Comparison of BIMA D-array and C$+$D-array maps of
3 fields.  BIMA catalog sources are shown by red dots, WS90
sources by blue dots.}
\label{fig:WScmp}
\end{figure*}

To examine a few of our clouds more closely, and to investigate
discrepancies between our catalog and the WS90 catalog, we obtained
higher resolution BIMA observations of the three fields shown by
dashed circles in Figure~\ref{fig:WS2}.  These fields were observed
on 3 nights in 2000 July with the array in its ``C'' configuration.  The
integration time was approximately 6 hours per field.  The C-array
observations were calibrated and combined with the D-array data to
generate maps with a $6.5''$ synthesized beam.  The RMS noise level in
each 2.03 \kms\ channel map was 0.15 Jy beam$^{-1}$, approximately
0.32~K, comparable to that of WS90.  The C$+$D-array maps are compared
with the D-array survey images in Figure~\ref{fig:WScmp}.

Generally there is good agreement between sources in our C+D array
images and the WS90 positions, although we fail to confirm some of the
less massive WS90 clouds (e.g., WS21, WS22, WS36).  One can find
several examples showing how the D-array survey blends WS sources
(e.g., WS27 and WS28 = catalog source no.  17; WS6 + WS8 = catalog
source no.  74).

\begin{figure}
\epsscale{1.0}
\plotone{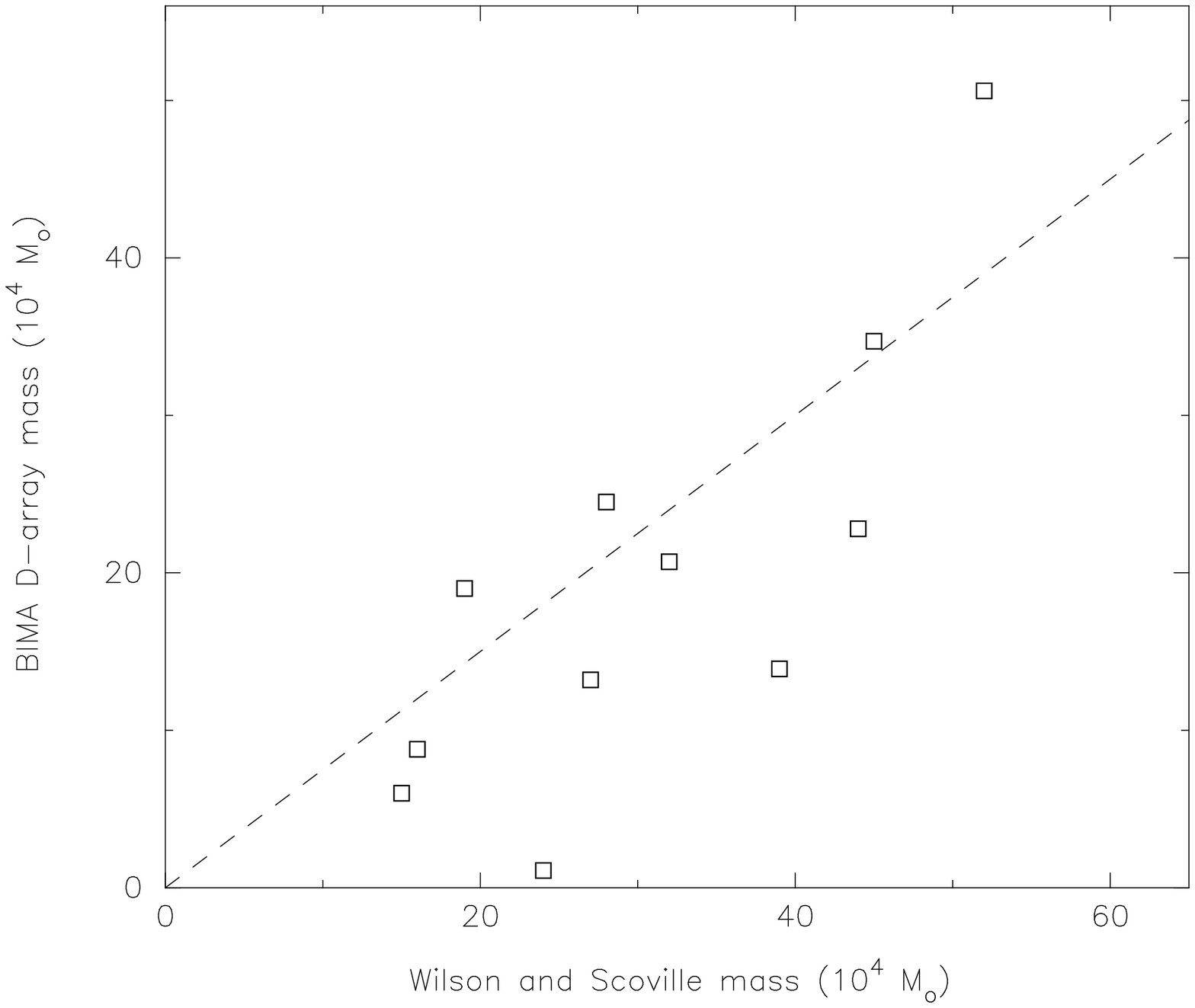}
\caption{Comparison of GMC masses from the BIMA D-array survey
and the WS90 survey, for cases where the same source clearly
is detected by both surveys.  The dotted line shows the 
expected slope, given the different mass conversion factors used
in the two surveys.}
\label{fig:WSmass2}
\end{figure}

Figure~\ref{fig:WSmass2} compares the masses of GMCs which can be
clearly identified in both data sets.  As indicated by the dashed line,
the BIMA masses are expected to be systematically lower by 25\% because
we used a lower conversion factor from flux density to mass.  Generally
the BIMA and WS90 masses are in reasonable agreement, although there is
a good deal of scatter in the ratio, probably because of the difficulty
in establishing the outer boundaries of clouds.

Small clouds detected by the C-array observations are not included in
our survey catalog (Table \ref{cloudprop}) but a further analysis of
these clouds is included in Paper II. 

\section{Cloud Properties}

\subsection{GMC Mass Spectrum}
\label{mass_spec}

\begin{figure}
\center
\plotone{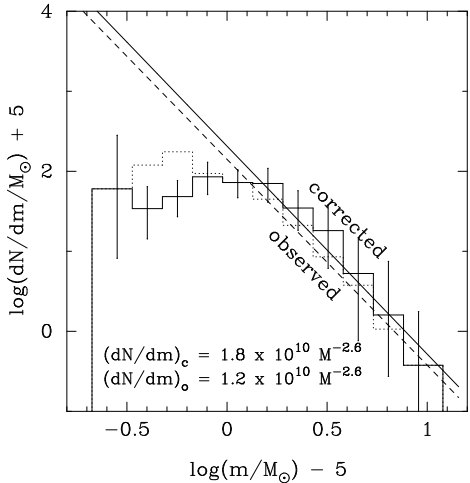}
\caption{The mass distribution $dN/dM$ of the GMC catalog.  Solid line
histogram shows the distribution of masses corrected for emission
profile truncation at 2$\sigma$. (Refer to
Figure~\ref{fig:frac_recovered}.)  Dotted line histogram shows the
distribution of uncorrected masses.  The histograms have been
accumulated into half-octave mass bins ($\Delta \log M = 0.5
\log_{10}2$), where the zero reference is $10^{5}~M_{\sun}$ on the
abscissa and $10^{-5}~M^{-1}_{\sun}$ on the ordinate. Power law fits
show that the mass correction has little effect on the slope of the
distribution for $M > 1.5\times 10^5~M_{\sun}$, but increases the
normalization coefficient 30\%.}
\label{fig:uspec}
\end{figure}

Histograms of the mass distribution of GMCs in our catalog are
presented on a log-log scale in Figure~\ref{fig:uspec}. The dashed
line histogram is the observed spectrum of GMC masses uncorrected for
signal beneath the $2\sigma$ clipping level, as described in \S
\ref{2sigma}. The solid line histogram is the spectrum of corrected
masses.  A power law of the form $$\frac{dN}{dM} = K \left(
\frac{M}{M_{\sun}} \right)^{-\alpha}$$ was fit to the mass bins for
which the catalog is complete ($M > 1.5 \times 10^{5}~M_{\sun}$). The
observed and corrected mass distributions are well fit with similar
indices ($\alpha = 2.6\pm 0.3$) but different normalization constants
($K=1.2$ vs. $1.8 \times 10^{10} M^{-1}_{\sun}$).

For a mass spectrum with $\alpha < 2$, most of the integrated mass
resides in the most massive clouds.  For a mass spectrum with $\alpha
> 2$, most of the mass is in the least massive clouds, and the total
mass diverges when extended to arbitrarily small mass. Hence, for M33
there must be either a lower mass limit for molecular clouds or a
break in the distribution where $\alpha$ changes with decreasing cloud
mass to less than 2.  If the integrated mass
\begin{equation}
M_{total}=\int_{low~limit}^{\infty} M~dN = \int_{low~limit}^{\infty} M
\left(\frac{dN}{dM}\right)dM
\end{equation}
is distributed according to  
\begin{equation}
\frac{dN}{dM} = 1.8 \times 10^{10} \left( \frac{M}{M_{\sun}}
~\right)^{-2.6}~[M^{-1}_{\sun}]
\end{equation}
we compute a strict lower mass limit $M_{L}$ of $4 \times
10^{4}~M_{\sun}$ for $\alpha = 2.6$ and $M_{total} = 4.5\times 
10^7\ M_{\odot}$.
More likely, there is a "turnover" mass, where the mass spectrum
becomes flatter than 2.
Since we can put a firm lower limit on the cutoff mass of $4 \times
10^{4}~M_{\sun}$ and an upper limit of $10^{5}~M_{\sun}$, where the
measured mass distribution is nearly complete, the turnover mass of
GMCs is $\sim~7 \times 10^{4}~M_{\sun}$ with an uncertainty of $3
\times 10^{4}~M_{\sun}$.  This turnover represents a
``characteristic'' mass for the molecular clouds in M33 that any
theory of GMC formation must be able to reproduce, given the physical
conditions in M33.


The highest mass we measure for a GMC in M33 is $7.3 \times
10^{5}~M_{\sun}$; there are two GMCs with this mass.  Neither is close
to the galactic center nor to an especially bright \ion{H}{2}~region.
In total, we find five GMCs more massive than the highest mass cloud
($4 \times 10^{5}~M_{\sun}$) reported by \citet{ws90}.  Moreover, they
find a lower value of $\alpha$ (1.7), which may have resulted from
preferentially targeting massive clouds by observing CO-bright single
dish fields.

\subsubsection{M33, Milky Way, and the LMC}             

Over a similar mass range, the power law distribution of M33 is
steeper than that of the Milky Way 
(1.6, Solomon et al. 1987; 1.9, Heyer, Carpenter, \& Snell 2001).
If M33 were as massive as the Milky Way, the normalization
constant of its power law distribution would be $3.6 \times
10^{11}~M^{-1}_{\sun}$ and $({dN}/{dM}) \Delta M \sim ~1.0$ at $1.0
\times 10^{6}~M_{\sun}$.  In contrast, the Milky Way is estimated to
have more than 100 GMCs with $M > 1.8 \times 10^6~M_{\sun}$
\cite[]{dame86}. We conclude that M33, even though it has many bright
\ion{H}{2} regions \citep{israel74}, has significantly fewer massive
GMCs.  As we infer in
$\S\ref{ha_compare}$, the majority of \ion{H}{2}~regions must be
(1) fueled by molecular clouds below our completeness threshold or
(2) capable of efficiently dispersing their parent GMC complexes
\citep{lh01}. Extrapolating the observed power law index to a cut-off
$3.0 \times 10^{4}~M_{\sun}$, we find $\int (dN/dM)\ dM ~\sim~2000$.
This is a sufficient number of molecular clouds to
match approximately the observed number of \ion{H}{2} regions.

The NANTEN survey of CO in the LMC revealed 168 GMCs \citep{fukui}.
The cloud sample has a mass range and completeness limit similar to
the BIMA survey. The mass spectrum of GMCs in the LMC has an index of
~1.9, similar to that of the outer Milky Way
\citep[]{hc01}, indicating that most of the mass resides in high mass
clouds. The physical meaning of the difference between the mass
indices of the LMC and M33 is unclear.  However, as in M33, molecular
clouds in the LMC appear to be tightly correlated with \HI\
overdensities \citep{mi01b}, and \HI\ filaments underlie bright
complexes of GMCs (\S\ref{GMC_HI}). Hence, we expect the details of
GMC formation to be at least somewhat similar for M33 and the LMC.
 
\subsubsection{Cloud Masses versus Radius} 

\begin{figure}
\center
\plotone{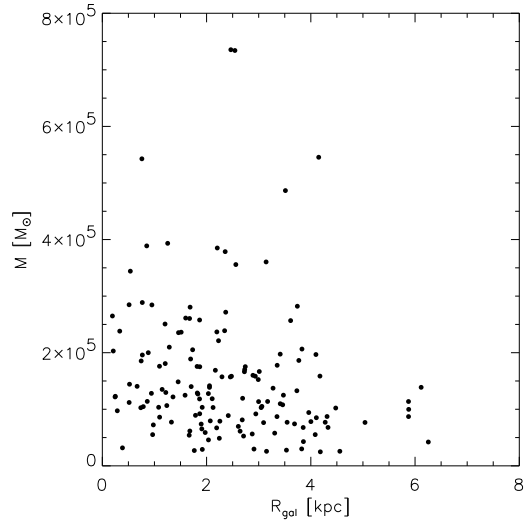}
\caption{
Catalog GMC mass ($M^*_{\mathrm{CO}}$) as a function of radius.  No
high mass clouds are found beyond 4.5 kpc. Within this boundary the mass
distribution is roughly independent of radius.  Survey coverage is
complete to 5.2 kpc and more than 50\% complete between 5.2 kpc and 8
kpc.  Accounting for incomplete coverage should, at most, double the
number of clouds seen in this range.
}
\label{fig:m_vs_r}
\end{figure}

Figure~\ref{fig:m_vs_r} shows the distribution of catalog GMC masses
as a function of galactocentric distance. 
The distribution of
clouds with masses greater than our completeness limit
(1.5$\times 10^{5}~M_{\sun}$) appears to cut off abruptly beyond 4
kpc. For $R_{gal} > 4.5$ kpc, we detect only a few low mass clouds
despite considerable reserves of \HI\ here (\S \ref{surf}).  However,
higher sensitivity measurements are required to determine if we are
indeed observing a sharp edge to the molecular disk or a shift to the
production of predominantly low mass clouds. For $R_{gal} < 4.5$ kpc,
the mass distribution appears more or less constant to the eye;
i.e. the apparent density of points in Figure~\ref{fig:m_vs_r} appears
constant with $R$ for a given mass.

In a more quantitative comparison, 
we looked for spatial variations in the GMC mass spectrum by
partitioning our catalog into an inner and outer disk sample, each
containing an equal mass of molecular gas in catalog GMCs. The radius
at which this occurs is 2.0 kpc. This coarse radial binning assures adequate
statistics for $M > 1.5 \times 10^{5}~M_{\sun}$ in two distinct cloud
populations.  Power law fits give $\alpha = 2.8\pm 0.3$ for the inner disk
and $\alpha = 2.4\pm 0.3$ for the outer disk.
A 2-sided K-S test shows that the two cloud samples are drawn from the same
population with a 9\% likelihood.  Though there is some evidence for a
difference in the mass distribution, it is only marginally significant. 

\section{Spatial Comparison of GMCs, \HI\ Structure, and \ion{H}{2} Regions}

\subsection{Radial Profiles and Star Formation Rates}
\label{surf}

\begin{figure}
\center
\plotone{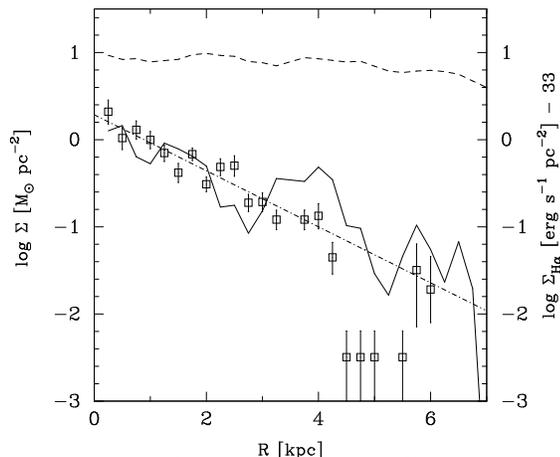}
\caption {The (deprojected) radial profiles of the molecular, atomic,
and ionized media.  All three are well fit by an exponential disk and
can be characterized by a peak value and a radial scale length.  Only
the fit to the molecular data is shown (dot-dash line).  Points
represent averages within deprojected radial bins with $\Delta R$ =
0.5 kpc.  The molecular data (squares) are well fit by an exponential
disk for $R < 4$ kpc with a scale length of $1.4\pm 0.1$ kpc and a
peak mass surface density, $\Sigma_{\mathrm{H2}} = 2.15\ M_{\sun}\
\mbox{pc}^{-2}$; $\Sigma_{\mathrm{H2}}$ drops sharply at $R=4$ kpc.
The \HI\ surface brightness (dashed line) is nearly constant for $R<6$
kpc with a scale length $10.6\pm 2$ kpc and a peak surface density of
$\Sigma_{\mathrm{HI}}(R=0)=9.5\ M_{\sun} \mbox{pc}^{-2}$ H$\alpha$
surface brightness (solid line) has a scale length of $1.7\pm 0.2$ kpc
and a peak surface brightness of $1.2 \times 10^{32} \mbox{ergs
s}^{-1}\mbox{ pc}^{-2}$.}  {\label{fig:surface_dens}}
\end{figure}

For galactic radii $R<7$ kpc, we derived the azimuthally averaged mass
surface density from our catalog GMCs. The results are presented in
Figure~\ref{fig:surface_dens}, along with the surface density of \HI\
mass and H$\alpha$ flux.  Points represent averages within deprojected
radial bins of $\Delta R$ = 0.25 kpc.  The \HI~data are from
\citet{cs00} and the H$\alpha$ data are derived from the optical
\ion{H}{2} region catalogs compiled by \citet{hw99}.  All three
components of the ISM are well fit by an exponential disk
$\Sigma(R)=\Sigma_0 \exp(-R/R_0)$.  Only the fit to the molecular data
is shown (dot-dash line) in the figure.  The molecular surface density
$\Sigma_{\mathrm{H2}}$ (squares) declines exponentially with $R_0 =
1.4\pm 0.1$ kpc, from a peak mass surface density $2.15\ M_{\sun}
\mbox{ pc}^{-2}$.  \citet{corb03} obtained a total molecular gas mass
of $2\times 10^{8}M_{\odot}$ from an FCRAO survey.  Accounting for the
difference in $X$ factors (\citet{corb03} used $2.8\times
10^{20}\mbox{cm}^{-2}/(\mbox{K~km~s}^{-1})$), her reported mass is 3
times the value we derive.  The surface mass densities in the central
region are comparable, but the scale length in \citet{corb03} is 2.5
kpc, resulting in a significantly larger integrated mass.  The reason
for the difference in the derived scale lengths is unclear and merits
further investigation.   The \HI\ surface brightness (dashed line) is nearly
flat for $R<6$ kpc with a scale length $10^{+8}_{-2}$ kpc and a peak
surface density of $9.5\ M_{\sun} \mbox{ pc}^{-2}$.  H$\alpha$ surface
brightness (solid line) has a scale length of $1.7\pm 0.2$ kpc and a
peak surface brightness of $1.2 \times 10^{32}~\mbox{ergs
s}^{-1}~\mbox{pc}^{-2}$, corresponding to a star formation rate of 9.5
$\times 10^{-10} M_{\sun}\ \mbox{yr}^{-1} \mbox{ pc}^{-2}$.

Kennicutt (1989,1998) has argued that there is a simple empirical
relation between the SFR and the surface density of gas,
$\Sigma_{gas}$, which can be parameterized by a power law SFR $\propto
(\Sigma_{gas})^{n}$. The SFR is most conveniently calculated by
scaling the H$\alpha$ line emission, powered by OB stars. Similar
scale lengths for H$\alpha$ and $\Sigma_{\mathrm{H2}}$ in M33 imply $n
= 0.9\pm 0.1$, consistent with a constant SFR per unit molecular mass.
Similarly, \citet{wb02} find for a sample of 7 normal spiral galaxies
that $\Sigma_{\mathrm{SFR}}$ strongly correlates with
$\Sigma_{\mathrm{H2}}$ only, where the power law index $n \sim$ 0.8
--- 1.4.  Integrating $L_{\mathrm{H \alpha}}$ for the catalog of
\citet{hw99} and using Equation 2 of \citet{k98}, we compute a star
formation rate of 0.24 $M_{\sun}~\mbox{yr}^{-1}$ and a molecular gas
depletion time of $1.9\times 10^8$ yr for M33.  This value is a factor
of 1.5 -- 4 smaller than the depletion times quoted in \citet*{wsr91}
who consider individual molecule-rich complexes as opposed to the
galaxy as a whole.  Our results, along with those of \citet{wb02},
are inconsistent with those of \citet{k89}, who finds for
disk-averaged observables that the SFR correlates more strongly with
the surface density of either \HI\ or \HI $+\mathrm{H_2}$ than with
that of molecular gas alone.

\subsection{Circular and Radial Motions of Molecular and Atomic Hydrogen}

We derived the rotation curve for GMCs cataloged
independently of $V_{\mathrm{HI}}$ (catalog sources 1 -- 93) and for the \HI\
emission along the same line-of-sight from the Westerbork data cube
\citep{deul87}.  The observed source positions $(x,y)$ were
deprojected onto the plane of the galaxy assuming a P.A. of $\sim
21^{\circ}$ and an inclination angle of $\sim 51^{\circ}$.  These
values were varied slightly with galactic radius in accordance with
the tilted ring model of \citet{cs00}.  The rotation center was
assumed to coincide with the optical center.  At the deprojected
positions the rotational velocity
\begin{equation}
v_{rot}(R) = \frac{V_{los}(x,y) - V_{sys}}{\sin i \cos \theta},
\end{equation}
where $\theta$ is the azimuthal angle of the source measured in the
plane of the galaxy from the the major axis.  Sources
within $45^{\circ}$ of the minor axis were
not included because of the large correction to ($V_{los}(x,y) -
V_{sys}$).

\begin{figure}
\center
\plotone{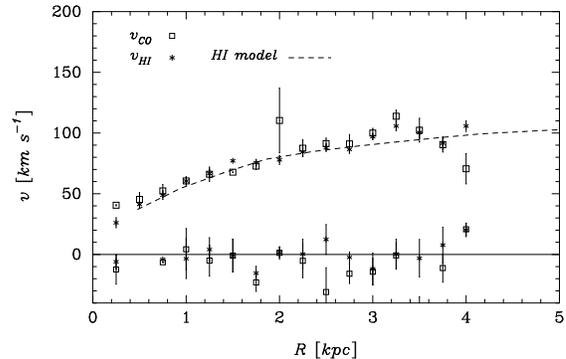}
\caption{ Radial profiles of circular and radial motion of primary
catalog GMCs, averaged over deprojected annuli of $\Delta R=250$
pc. Model rotation curve of \HI\ also shown (dash). To measure the
radial motions, only clouds within $30^{\circ}$ of the minor axis were
included.  }
\label{fig:Vco_R2}
\end{figure}

Figure~\ref{fig:Vco_R2} shows the circular velocities for catalog
GMCs, averaged in 250~pc bins.  The median CO rotation curve (open
squares) closely matches the \HI\ rotation curve determined by
\citet{cs00} from Westerbork data.  The dispersion of GMCs about the
\HI\ rotation curve varies between 7.5 and 9.0 \kms~ for $R < 4~$
kpc. There is no apparent trend in the dispersion with galactic
radius.  These values are comparable to Milky Way values after
streaming in the Galaxy is taken into account.

Figure~\ref{fig:Vco_R2} also shows the CO radial velocities
\begin{equation}
v_{rad}(R) = \frac{V_{los}(x,y) - V_{sys}}{\sin i \sin \theta},
\end{equation}
including only sources within $\Delta \theta < 30^{\circ}$ of the
minor axis.  We find little evidence for large scale radial
(non-circular) motions in M33 in agreement with \citet{deul87}.

\subsection{Spatial Comparison of GMC Distribution with \HI\ Filaments}
\label{GMC_HI}

\begin{figure*}
\begin{center}
\plotone{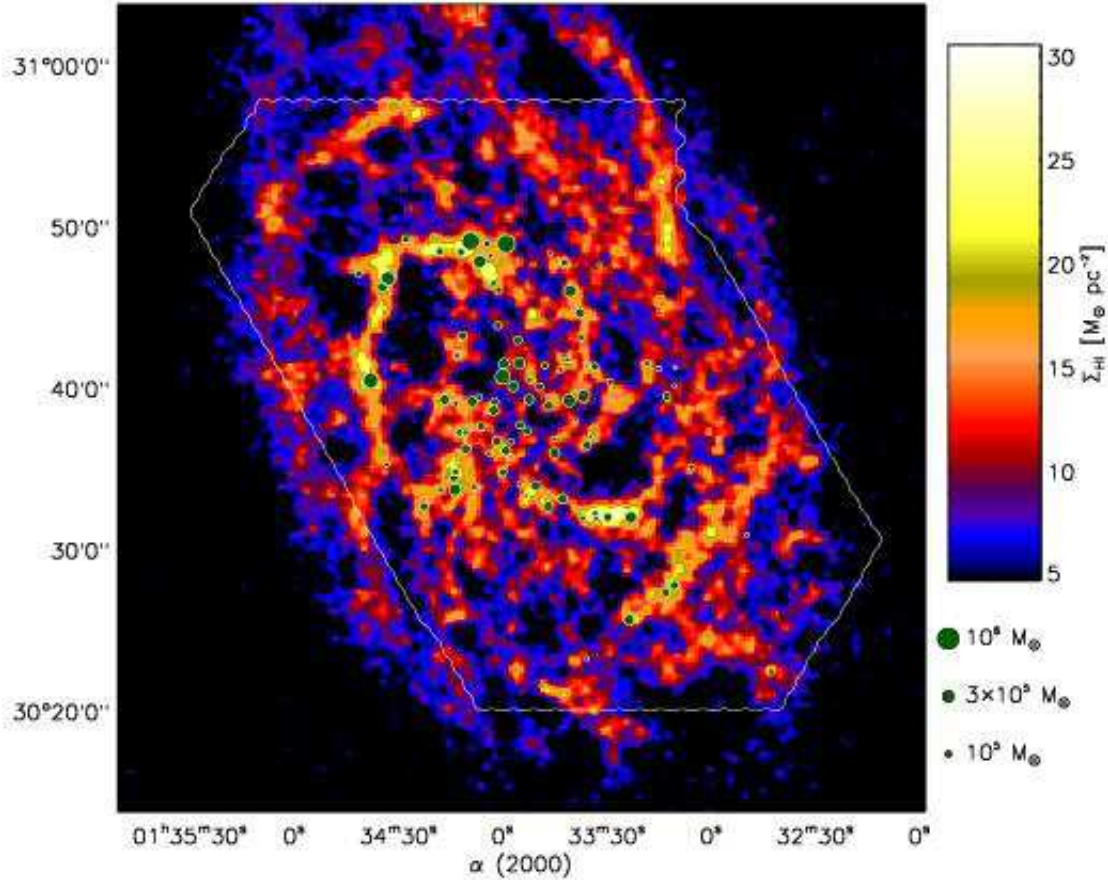}
\caption{ Color image of
\ion{H}{1} 21 cm emission from \citet{deul87}  with
catalog sources overlayed.  All molecular clouds lie in regions of
\ion{H}{1} overdensity.  The area of the molecular cloud has been
scaled to represent the relative masses of the clouds.  The
coincidence of molecular clouds with \ion{H}{1} overdensity is
evidence that clouds form out of the atomic gas. }
\label{fig:h1map1}
\end{center}
\end{figure*}

\begin{figure}
\begin{center}
\plotone{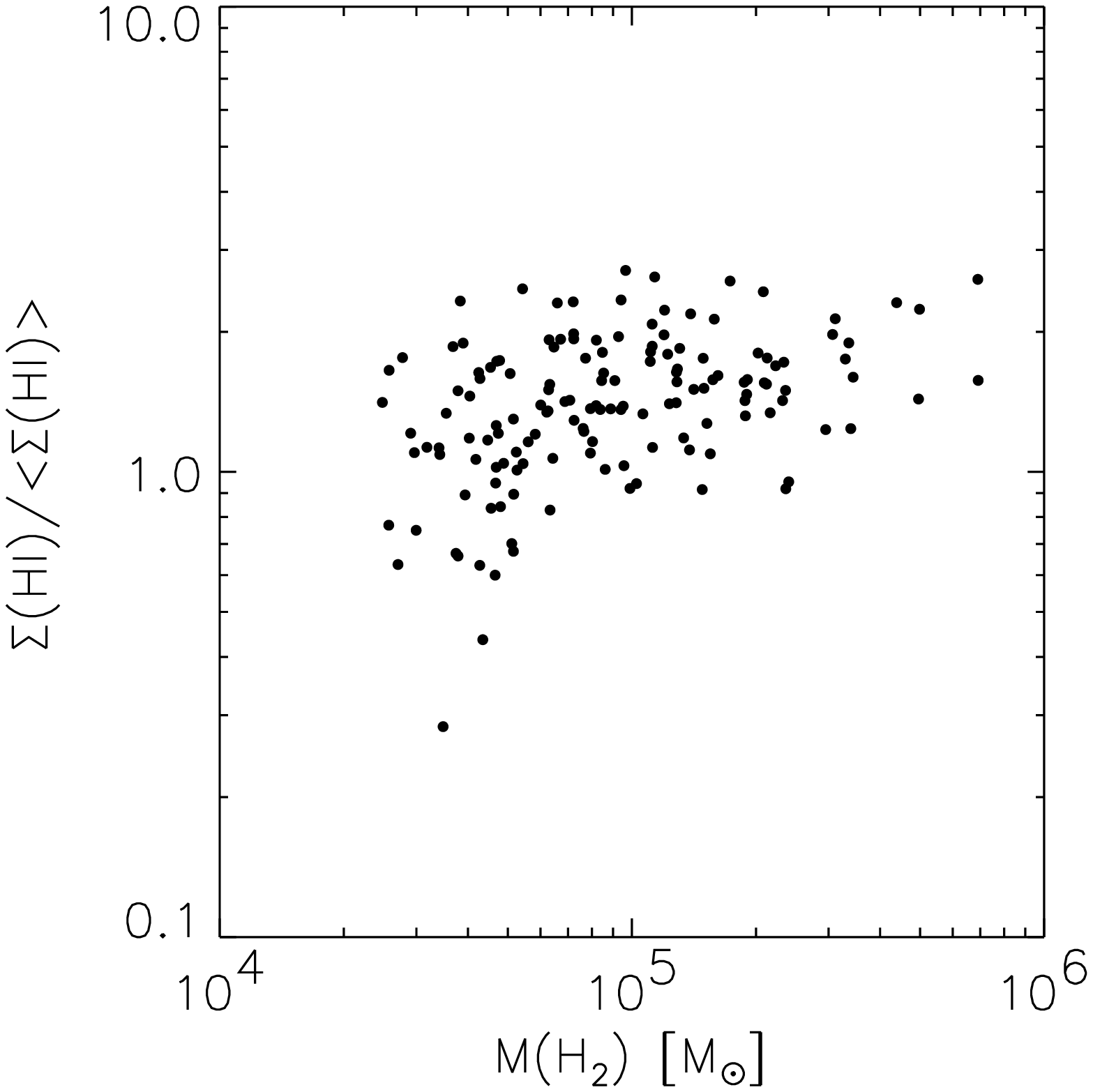}
\caption{Plot of the mass of a GMC against the \HI\ overdensity at the
galactic radius of the cloud.  The overdensity is defined as the \HI\
surface density observed at the position of the cloud divided by the
azimuthally averaged \HI\ surface density found at that radius.  The
plot shows that there is no relation between these two quantities.
Note that a few low mass clouds are found in regions of \HI\
underdensities.  These may be among the clouds in the catalog that are
not real.}
\label{fig:h2mass}
\end{center}
\end{figure}

An extraordinary spatial correspondence between GMCs and the
distribution of atomic hydrogen is shown in Figure~\ref{fig:h1map1}.
The halftone image is the \ion{H}{1}~ brightness distribution, derived
by integrating the Westerbork data cube \citep{deul87}; the resolution
of the Westerbork map is $24'' \times 24''$, comparable to the BIMA
data. The \ion{H}{1}~ distribution is dominated by bright, narrow
filaments
in an extended disk. The dark circles mark the positions of the
catalog GMCs; cloud masses are proportional to circle areas.
Figure~\ref{fig:h2mass} shows that nearly all GMCs are found in
regions of \HI\ overdensity, but there is no correlation between the
magnitude of the overdensity and the cloud mass.  Apparently, the
amount of ambient \ion{H}{1} is independent of the mass of the GMC
that forms from it.  The correspondence between GMCs and filaments
persists down to smaller scales in the central region where the
\ion{H}{1} brightness map has less dynamic range. The association with
the filaments is even more pronounced in the high resolution \HI\ maps
of M33 made at the VLA \citep{thilker}.  The average \HI\ mass located
within 150 pc of our catalog sources is $5\times 10^{5}~M_{\sun}$,
indicating an ample local supply of atomic hydrogen from which
observed GMCs could have condensed.

\begin{figure}
\begin{center}
\plotone{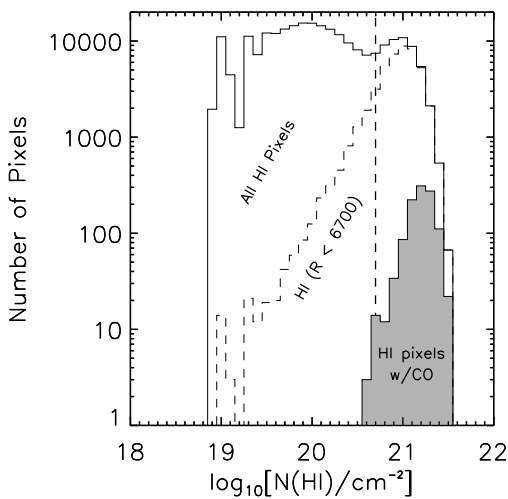}
\caption{Histogram of \ion{H}{1} column densities in the WSRT map of
M33 \citep{deul87}.  The solid histogram shows the column densities
for pixels through the entire map.  The dashed histogram represents
pixels within 6.7 kpc of the center (the H$\alpha$ disk).  The shaded
histogram represents pixels associated with CO emission, that is
pixels within the 2$\sigma$ boundaries of our 148 cataloged
clouds. The dotted vertical line at $5 \times 10^{20}\mbox{ cm}^{-2}$
is the threshold column density for molecule formation in the solar
vicinity \citep{savage}.}
\label{fig:h1cohist}
\end{center}
\end{figure}

In Figure~\ref{fig:h1cohist}, we see that GMCs form from \HI\ with an
column density threshold of $\sim5 \times 10^{20}~\mbox{cm}^{-2}$,
which is comparable to the mean \HI\ column density in the H$\alpha$
disk (dashed histogram).  The shaded histogram indicates pixels with
associated CO emission.  The solid line shows all WSRT pixels
\citet{deul87}.  The threshold value is similar to that found by
\cite{savage} and \cite{bdd} in the solar vicinity, and is indicated
in Figure~\ref{fig:h1cohist}.  Moreover, the \HI\ shows an upper limit
in column density of $\sim 3\times 10^{21}\mbox{ cm}^{-2},~ A_V\sim
3$, often at positions associated with molecular gas.  This suggests
that the atomic gas is providing shielding for the molecular gas, but
higher column densities result in more gas in the molecular state.

{\it The close association between GMCs and \HI\ filaments can be used
to set an upper limit on the GMC lifetimes.}  The RMS difference
between the CO and \HI\ velocity centroids is 8 \kms, calculated from
clouds 1--93, which were selected independent of the \HI\ velocity.
The typical width of an \HI\ filament in the \citet{deul87} map is
$\sim 200$ pc.  If cloud lifetimes were significantly longer than
10--20 Myr, clouds would drift off the filaments, destroying the
observed association.

\subsection{Implications for GMC Formation}

\citet{eg90} reviews many theories of GMC formation, including the
agglomeration of small molecular clouds \citep{sh79}.  However, there
can be little doubt that the GMCs in M33 form out of the atomic gas.
If instead the atomic gas resulted from the dissociation of molecular
gas, the \HI\ would appear as discrete envelopes around the GMCs
rather than as filaments, and there would little atomic gas beyond the
apparent edge of the molecular disk at $R=4$~kpc. Neither of these
features are seen.  In addition, there is significantly more atomic
gas ($\sim$ a factor of 5) in the filaments than can be associated
with discrete envelopes around GMCs and $<15\%$ of this mass can be
attributed to photodissociation \citep{ws91}.  Furthermore, if GMCs
formed by the collisional agglomeration of smaller {\it molecular}
clouds, the growth time would be comparable to the collision time.
This is roughly equal to the collision time of the catalog clouds,
which is $2 \times 10^9$ years based on 140 GMCs inside a radius of 4
kpc, an H$_2$ half thickness of 100 pc and a cloud-cloud velocity
dispersion of 10 \kms.  Since the number density of clouds scales
roughly as $1/M_{cloud}$ and the geometric cross section scales as
$M_{cloud}$ (owing to GMCs having a constant surface density, see
Paper II), this collision time is independent of mass and
characterizes the growth time for all self-gravitating clouds.
Gravitation effects do not significantly enhance the collision cross
section \citep{sh79,bs80}.  Additionally, Paper II sets an upper limit
for the diffuse H$_2$ surface density of 0.3 M$_\sun$ pc$^{-2}$.
Forming a $10^5$ $M_\sun$ GMC from a diffuse molecular component at
this surface density would require almost 10$^8$ yr to accumulate
material from a 300 pc radius region, provided that the gas velocity
dispersion could be converted into ordered motion.  Both times are
considerably longer than the lifetimes of the GMCs ($\sim 2 \times
10^7$ yr) based on their association with the \HI\ filaments.

It appears that the process of GMC formation in M33 is first and
foremost a process of \HI\ filament formation.  While a high \HI\
column density is a necessary condition for the formation of a GMC, it
clearly is not a sufficient condition, since there are many positions
with high \HI\ column densities but no observed CO emission
(Figure~\ref{fig:h1cohist}).  The filamentary structures extend out to
the $R = 8$~kpc edge of the \HI\ map, well beyond the radius where
GMCs are observed, with no significant change in character at the edge
of the molecular disk.  It appears that some physical parameter, which
evidently is a function of galactic radius, determines what fraction
of the atomic gas in a filament is converted into molecules.  A likely
candidate is hydrostatic pressure in the disk, a hypothesis that will
be investigated in a subsequent paper.
 
\subsection{Are GMCs and \HI\ Holes Physically Linked?}
\label{holes}

Expanding shells in the interstellar medium may play an important role
in the dynamics of the neutral gas, transmitting both mechanical and
radiant energy at their boundaries \citep{mo77}.  \citet[][
DdH]{DdH90} cataloged 148 holes (underdensities) in the \HI\
distribution of M33. Ninety-three of the holes (Types 2 and 3 in DdH)
show the kinematic evidence of expansion in the Westerbork channel
maps.  These holes cover $< 20\%$ of the neutral disk area.  DdH find
that many of the compact holes have OB associations inside and
conclude that many such holes have been excavated by massive star
formation activity.  While DdH also allow for the possibility that
holes are formed by the local collapse of \HI\ to form GMCs, our CO
survey shows that this is unlikely, since nearly all GMCs are {\it
outside} the holes.  In this section we consider evidence of a
physical connection between GMCs and the \HI\ shells.

\begin{figure}
\plotone{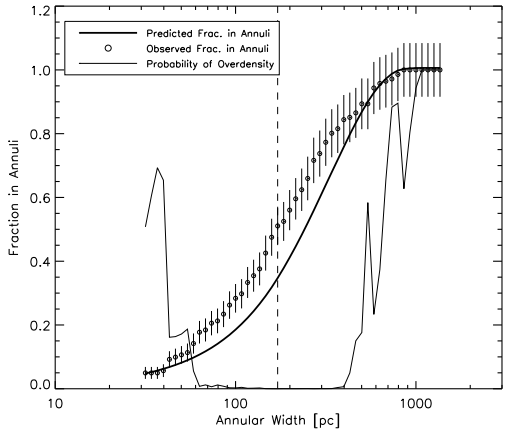}
\caption{ Positional correlation of catalog GMCs with ~\ion{H}{1}
holes cataloged by \citet{DdH90}. Open circle denote the fraction of
clouds which overlap an annulus of width $\Delta R$ bounding some hole with
diameter $<$ 400 pc.  The thick curve denotes the fraction of overlap
expected for randomly distributed clouds. The thin curve is the
binomial probability for the observed deviation, which is $<$ 0.05 for
annular widths of 60 --- 300 pc. The minimum probability 0.003, marked
by the dashed line, occurs at an annular width of 150 pc. 
}
\label{fig:hole_corr}
\end{figure}

\begin{figure}
\plotone{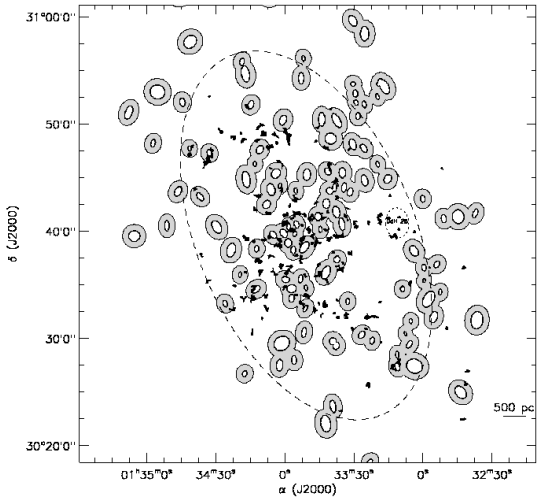}
\caption{ HI holes and associated GMCs.  Shown are 2$\sigma$
boundaries (filled) of catalog GMCs and 150 pc annuli (shaded areas)
around DdH hole edges.  From this figure it can be seen that no
significant CO emission extends into the DdH holes.  The large
separation (150 pc) between hole edges and GMCs suggests that
expanding shells do not {\it directly} trigger formation of the
catalog GMCs. An extraordinary hole-GMC association (labeled DdH 28)
may be evidence of a high velocity cloud impacting the \HI\ disk.  }
\label{fig:donut}
\end{figure}

There is, in fact, a tendency for the catalog GMCs to be spatially
clustered around the DdH \ion{H}{1} holes.  We quantified this effect
by measuring the fraction of GMCs located within annuli of width
$\Delta R$ bounding the holes, and comparing this to the fraction
expected if the GMCs and holes are uncorrelated. We restricted the
analysis to holes with diameters less than 400 pc because these are
more likely to have been created by high mass star formation.  The
results are shown in Figure~\ref{fig:hole_corr}.
We find a significant correlation between GMCs and \HI\ hole edges for
$\Delta R$ between 60 and 300 pc, most significantly for $\Delta R$ =
150~pc (binomial probability $P_{rand}\sim 10^{-3}$).  The large
separation (150 pc) between hole edges and GMCs suggests, however,
that expanding shells do not {\it directly} trigger formation of the
GMCs.  Figure~\ref{fig:donut} plots the GMCs (boundaries defined by
the 2$\sigma$ clipping levels) and the 150 pc annuli (shaded) around
the (compact) hole edges.  This figure shows that there is no
significant CO emission inside the DdH holes.

Since essentially all GMCs are on filaments but not all GMCs are at
the edges of holes, we conclude that the filamentary structure is more
fundamental to GMC formation than the action of expanding shells.
These shells may create filamentary structure on small scales,
especially in the presence of turbulence \citep[e.g.][]{lh01}, but are
unlikely to produce the largest filaments.  The energies required to
evacuate the large holes defining these filaments are $\geq 10^{53}$
ergs (see e.g. DdH).  Such energies would leave bright stellar
remnants at the centers of the large holes, which are not generally
observed.  In addition, these holes would be sheared apart on time
scales of $2 \times 10^7$ years by differential galactic rotation
unless they were stabilized by some process.  Thus some organizing
principle other than stellar feedback seems to be responsible for much
of the filamentary \HI\ structure, and ultimately, for GMC formation
in M33.  One possibility is MHD instabilities in a self-gravitating,
shearing disk \citep{eg87,kos02}.

Although nearly all GMCs are outside holes on filaments of \HI , an 
exception is DdH 28 (Figure \ref{fig:donut}).  The catalog GMCs in
this hole may result from a high velocity cloud (HVC) colliding with
the \ion{H}{1} disk. This scenario for hole creation was first elaborated
by \citet{TT81}. GMCs 66 and 88 have velocities blue-shifted relative
to local \HI\ by 72 and 24 \kms, respectively.  The formation of GMCs
40, 51, 54, 123, and 133 at the edge of the hole -- all blueshifted
relative to local \HI --  may have been triggered by this collision as well.
The combined kinetic energy of these clouds ($1.4\times 10^{52}$ ergs)
could have been supplied by an atomic HVC. Alternatively, GMCs in DdH
28 may have formed on the front surface of neutral shells. That is,
they may be similar to the Orion A and B molecular clouds, which appear
to have formed from gas lifted 140 pc out of the plane of the Milky Way
by stellar winds \citep{dame01, lh01}.

\subsection{Spatial Comparison of GMC Distribution with \ion{H}{2}\ Regions}
\label{ha_compare}

\begin{figure*}
\center
\plotone{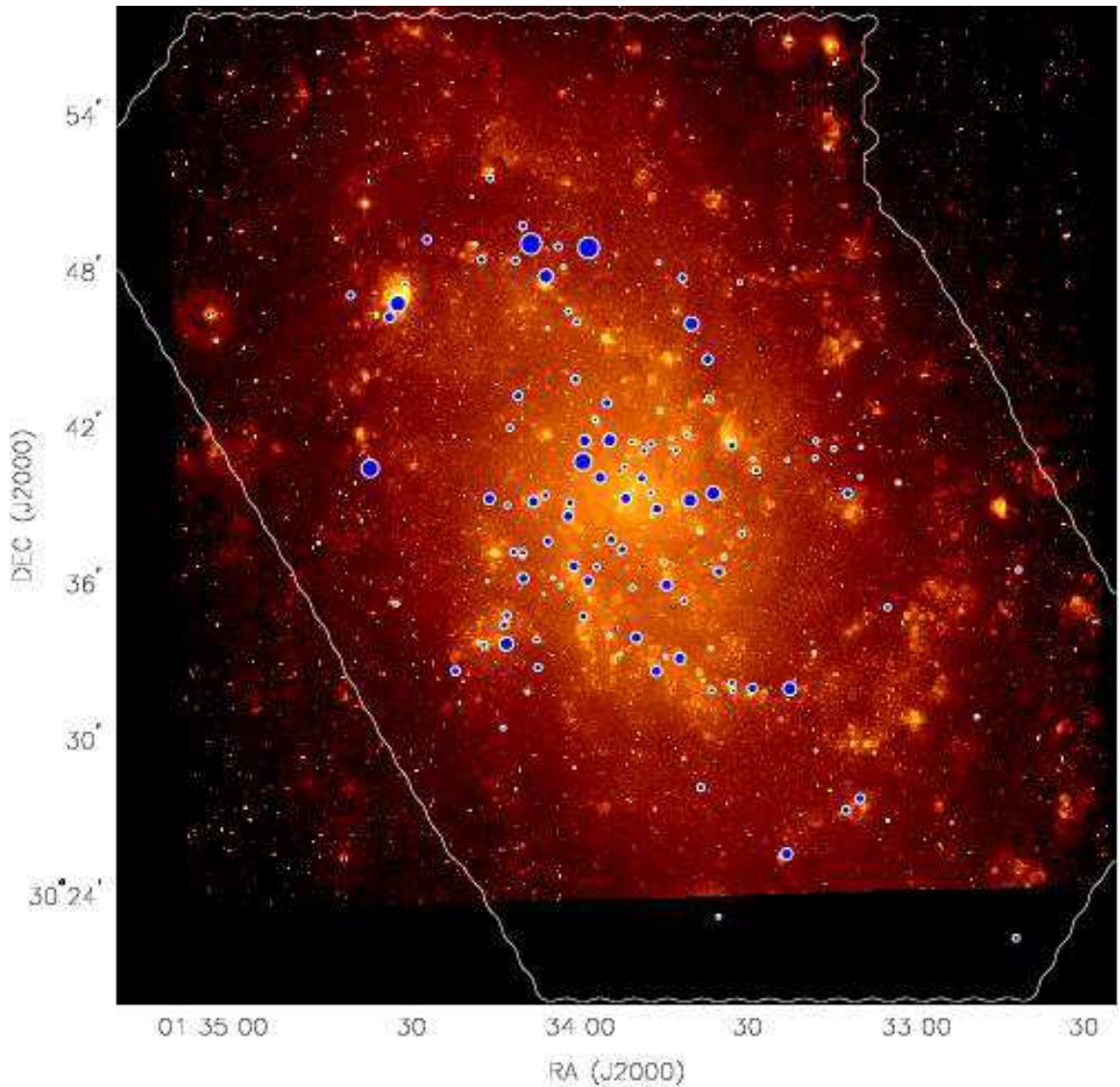}
\caption{ Location of GMCs plotted with blue circles on the \ha~image
of \citet{Cheng96}.  The area of circles is proportional to the
molecular mass.}
\label{fig:CO-on-Ha}
\end{figure*}

Figure~\ref{fig:CO-on-Ha} compares the locations of GMCs with
H$\alpha$ emission, showing that some GMCs follow the weak \ha~spiral
arms of M33.  Molecular clouds in the overlay are represented by
filled circles with area proportional to mass; the underlying halftone
is a CCD image from \citet{Cheng96}.  Most of the \ha\ flux comes from
\HII\ regions, which are cataloged in \citet{bHa74}, \citet{wh97}, and
\citet{hw99}.  As shown in \S\ref{surf}, the radial profiles of CO
flux and \HII\ region flux have comparable scale lengths; however, not
all \HII\ regions are associated with GMCs. In fact, there are over 20
times as many \HII\ regions as catalog GMCs, and many of the \HII\
regions are far from detected clouds.  Since our GMC catalog has a
lower dynamic range than the \HII\ region catalog, it's possible that
low mass GMCs are associated with most of these ``orphan'' \HII\
regions; \S\ref{mass_spec} argues that there could be $\sim$ 2000
molecular clouds in M33, comparable to the number of \HII\ regions.
Alternatively, \HII\ regions may rapidly dissipate molecular clouds,
accounting for the discrepancy in numbers \citep{lh01}.

We calculate the fraction of the H$\alpha$ flux that originates from
\HII\ regions that spatially overlap catalog GMCs.  Using the
positions and radii of \ion{H}{2} regions from the combined catalogs
\citep{bHa74,wh97,hw99}, we find that 40\% of the \ha\ flux originates
from \HII\ regions which are tangent to, or contain, catalog GMCs.  If
the remaining $2.2 \times 10^7\ M_{\odot}$ of molecular gas
(\S\ref{sens_budget}) produced a proportional amount of \ha\ flux, at
least 80\% of the emission would arise from \HII\ regions associated
with GMCs.

\begin{figure}
\center
\plotone{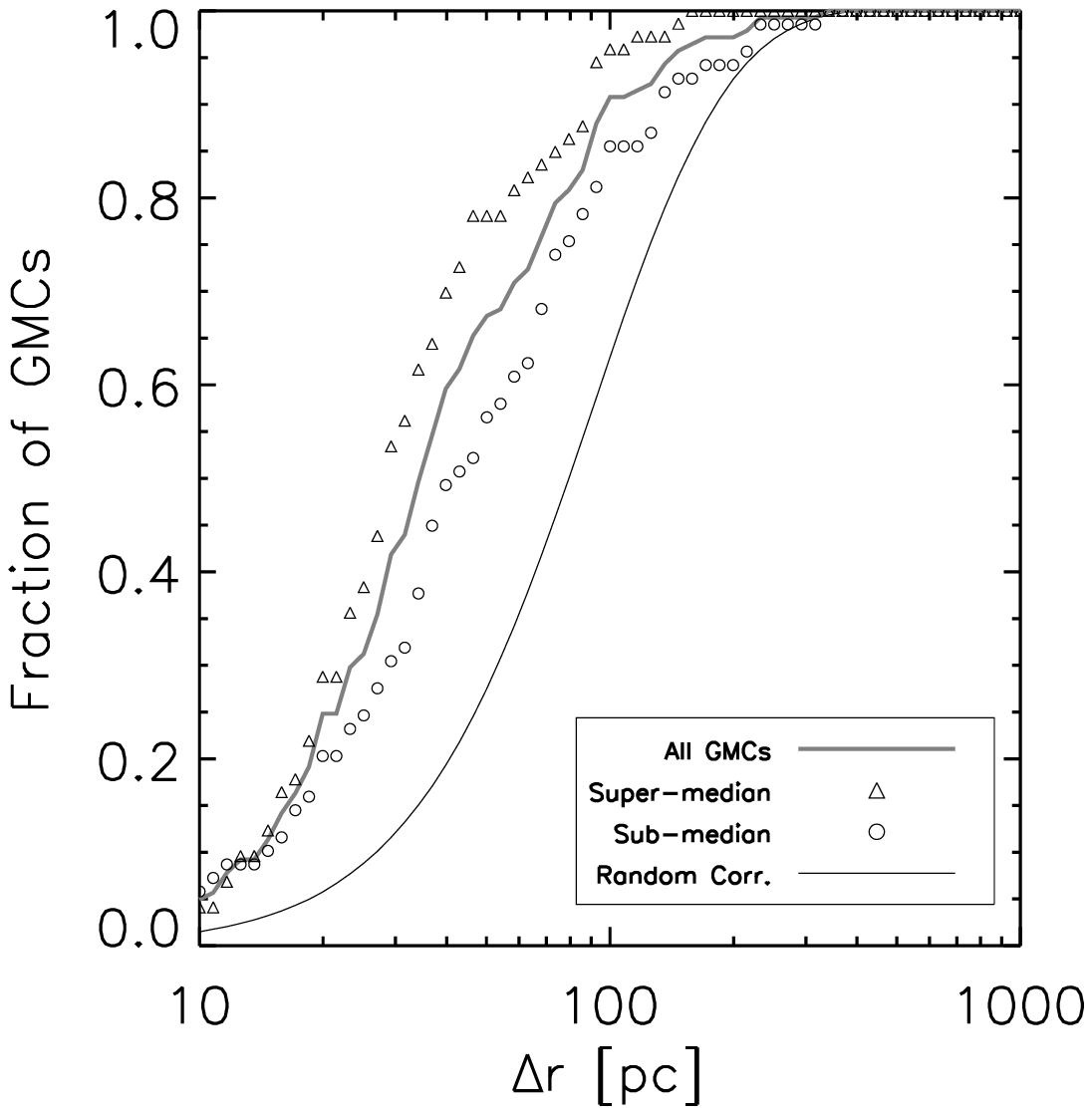}
\caption{
Figure~\ref{fig:h2corr} shows the positional correlation of catalog
GMCs with \ion{H}{2}~regions.  The \ion{H}{2}~region
positions are from the combined catalogs of \citet{bHa74},
\citet{wh97}, and \citet{hw99}.  We measured the fraction of GMCs with
at least one \ion{H}{2} region within a separation $\Delta r$.
The thin curve is
the fraction expected from random association.
There is statistically significant clustering of GMCs and
\ion{H}{2}~regions out to a separation of 150 pc.
As many as 100 GMCs (67~\%) are apparently forming
massive stars -- that is, their centroid position is within 50 pc of
some \ion{H}{2} region.
85\% of the higher mass GMCs (triangles) are within 50 pc of \ion{H}{2}
regions versus 55\% of the lower mass GMCs (circles).  This result is consistent
with a shorter quiescent {\it time} for higher mass
clouds.
}
\label{fig:h2corr}
\end{figure}

One may also ask what fraction of the GMCs are actively forming stars.
Figure~\ref{fig:h2corr} shows the positional correlation of catalog
GMCs with \ion{H}{2}~regions (circles).  We counted the fraction of
GMCs that have at least one \ion{H}{2} region within distance $\Delta
r$.  The thin curve is the fraction expected from random association
(see Appendix B).  There is a statistically significant clustering of
GMCs and \ion{H}{2}~regions out to a separation of 150 pc.  As many as
100 GMCs (67\%) are apparently forming massive stars -- that is, their
centroid position is within 50 pc of some \ion{H}{2} region.  This is
a similar fraction of association as found in a more limited study by
\citet{ws91}.  Of the remaining GMCs, we estimate that only a few
contain obscured H$\alpha$ sources.  From our correlation curve, there
are only 7 \ion{H}{2} regions coincident with a GMC position ($\Delta
r \leq 10$ pc). By symmetry, we expect that only $\sim 7$ catalog GMCs
are obscuring signs of massive star formation.

Winds and ionizing radiation from massive stars destroy molecular
clouds.  Since at least 2/3 of our sources are associated with \HII\
regions, we infer that GMCs spend less than 1/3 of their life in a
quiescent phase prior to the onset of massive star formation.  The
fraction of time spent in the quiescent phase appears to be even
smaller for the more massive clouds. Figure \ref{fig:h2corr} shows
that 85\% of the higher mass GMCs are within 50 pc of \ion{H}{2}
regions versus 55\% of the lower mass GMCs.  A reciprocal correlation
exists between luminous \HII\ regions and GMCs: luminous \HII\ regions
(top 10\%) are nearly twice as likely to have an associated catalog
GMC than the population as a whole.

Between the cutoff of the molecular disk ($R=4.5$ kpc, \S\ref{surf})
and the edge of the star forming disk ($R=6.7$ kpc), there are 7
cataloged GMCs and 600 \HII\ regions.  Comparing the total \ha\
luminosity in this region to the total cataloged molecular mass, we
find the ratio $L_{\mathrm{H}\alpha}/\sum M_{\mathrm{H2}}$ for this
region is 8 times larger than in the inner galaxy ($R < 4.5$ kpc).
Correcting for different sensitivities and survey coverage in the two
regions can only account for a factor of two discrepancy.  Based on
the H$\alpha$ flux, we would expect to have cataloged {\it at least} 4
times as much molecular mass in the outer galaxy as we actually do.
It is unlikely that a burst of star formation in this region has
depleted the supply of GMCs.  We conclude either that dispersal times
for GMCs are significantly shorter in the outer galaxy, that the mass
spectrum of the GMCs becomes even steeper there, or the metallicity
significantly affects the CO-to-H$_2$ conversion factor in this
region.

\section{Summary}

In this paper we present the results of an unbiased interferometric
survey of the star forming disk of M33 in CO(1$\rightarrow$0); the
observations were done using the BIMA array.  The 50 pc linear
resolution of our survey map is comparable to the size of most GMCs.
We derive a catalog of GMCs for M33 complete down to 1.5 $\times
10^{5}~M_{\sun}$.  From simple statistics, we expect that no more than
15 of the 148 sources listed in our catalog are spurious.  We estimate
that approximately 60 clouds with a total mass of $4.3\times
10^6~M_{\odot}$ have been falsely rejected from the catalog.  The
interferometer data were compared with single dish fields observed at
the UASO 12 m telescope to estimate the effects of spatial filtering
and masking on the catalog.  Our results generally agree with those of
\citet{ws90}, but the greater spatial coverage and flux completeness
of the BIMA survey leads, in some cases, to different physical
inferences.

From the survey data, we conclude the following:   

1. The total mass of GMCs in our catalog is $2.3\times
10^7~M_{\odot}$.  From a comparison of the interferometer and single
dish data in selected fields, we estimate that the total molecular
mass of the galaxy is $4.5\times 10^{7}~M_{\odot}$, 2\% of the atomic
mass.

2. GMCs with masses above our completeness limit are described by a
power-law distribution with $dN/dM \propto M^{-2.6\pm 0.3}$.  The power-law
index of GMCs in M33 is steeper than that in the Milky Way ($dN/dM
\propto M^{-1.6}$).  We infer that there exists a low mass cutoff or a
change in the index to $<$ 2 between $3\times 10^{4}~M_{\odot}$ and
$10^5~M_{\odot}$.  The cutoff or change in slope implies that GMCs in
M33 form with a characteristic mass of $\sim 7 \times
10^{4}~M_{\odot}$.  

3. The surface density of molecular gas decreases exponentially in 
radius, with a scale length of $1.4 \pm 0.1$ kpc.  The
scale length agrees well with the surface brightness of H$\alpha$
emission ($1.7 \pm 0.2$ kpc) implying a scaling between
star formation and molecular gas surface density of SFR $\propto
\Sigma_{\mathrm{CO}}^{0.9\pm 0.1}$.  Based on the H$\alpha$ luminosity
of M33, we estimate a star formation rate of 0.24 $M_{\odot}$
yr$^{-1}$, implying a molecular gas depletion time of $1.9 \times
10^{8}$ yr.  This very short time is not unreasonable given the large
reservoir of atomic gas available at all radii.

4. The rotation curve of the galaxy derived from CO emission is in
excellent agreement with that of the \ion{H}{1}.  We find no evidence
for large scale radial motions of molecular gas in the galaxy.

5. Giant molecular clouds are preferentially found on bright
\ion{H}{1} filaments where the \ion{H}{1} column is approximately
$10^{21}$ cm$^{-2}$.  Regions with \HI\ column densities $<5 \times
10^{20}$ cm$^{-2}$ are devoid of catalog GMCs.
 
6.  We estimate a GMC lifetime of 10--20 Myr based on the close
association between GMCs and \HI\ filaments.  Given the typical
velocity difference of 8 \kms, the clouds would drift away from the
filaments on longer time scales.

7.  Since the filamentary structure of the \HI\ gas appears to be 
independent of the existence of nearby GMCs, we conclude
that it is the \HI\ filamentary structure that forms first.  The
fraction of gas that becomes molecular is then determined by some
other parameter, such as hydrostatic pressure, that is inversely
correlated with galactic radius.
  
8. At least 40\% of the H$\alpha$ flux is associated with catalog
GMCs.  If the undetected molecular mass produces a proportional amount
of H$\alpha$ flux, at least 80\% of the emission would arise from
\HII\ regions associated with GMCs.  At least 2/3 of catalog GMCs are
within 50 pc of an \ion{H}{2} region, so it appears that GMCs spend
less than 1/3 of their lifetime in a quiescent phase prior to the
onset of star formation. Clouds above the median mass spend less than
15\% of their lifetime in the quiescent phase.  \HII\ regions above
the 90th percentile in luminosity are nearly twice as likely to have a
GMC within 50 pc as compared to the entire population.

9. Nearly all catalog GMCs are exterior to \HI\ holes.  Clearly, \HI\
holes do not result from conversion of atomic to molecular gas.  GMCs
and compact \HI\ holes ( $d < 400$ pc) are clustered over separation
scales of 60 --- 300 pc.  However, since many GMCs are not associated
with holes, the holes probably play only a secondary role in GMC
formation.


10. GMCs 66 \& 88 are the only GMCs interior to an \HI\ hole (DdH 28);
they are highly blueshifted with respect to \HI.  GMCs 40, 51, 54, 123
\& 133, located along the edge of DdH 28, are also blueshifted with
respect to \HI.  Conceivably these clouds were formed by the impact of
a high velocity cloud on the atomic disk.

11.  Although the molecular disk shows a sharp decline in the number
of GMCs beyond $R \sim 4.5$ kpc, the H$\alpha$ disk extends to $R \sim
6.7$ kpc.  This suggests that, for the outer disk, either dispersal
times of GMCs are significantly shorter than for the inner disk or
that there exists a large population of GMCs below our detection
threshold.


\acknowledgements

This work was partially supported by NSF grant AST-9981308 to the
University of California. This research made extensive use of NASA's
Astrophysics Data System (ADS) and the NASA/IPAC Extragalactic
Database (NED). ER's work is supported in part by a NSF Graduate
Fellowship.


\clearpage

\appendix

\section{Probability calculation}

This appendix provides a brief description of the calculation used to
rank the likelihood of candidate molecular clouds in the data cube. 

As discussed in the text, the flux density $S$ at each pixel in the data
cube is first normalized by the RMS noise $\sigma$, as a function of
position and velocity, to produce a signal-to-noise, or significance,
image $y$: 
\begin{equation}
s(x,y,v) = \frac{S(x,y,v)}{\sigma(x,y)\sigma(v)}.
\end{equation}

Suppose that at some location in the data cube we have identified a
spectral feature, occupying $n$ adjacent velocity channels, which we
suspect may be CO emission from a molecular cloud.  What is the
likelihood that this is a false detection?

Assume that the signal-to-noise ratios of the $n$ channels have been
sorted in ascending order such that $s_1 \leq s_2 \leq \ldots \leq
s_n$.  We wish to calculate the probability of randomly drawing $n$
values from a normalized Gaussian distribution (with zero mean and
standard deviation 1), such that all $n$ values are $\geq s_1$, at
least $(n-1)$ values are $\geq s_2$, \ldots, and at least 1 value is
$\geq s_n$ -- that is, the probability of selecting $n$ values which
are at least as unlikely as the actual spectrum.

Suppose that one draws a single value $s$ from the Gaussian
distribution.  There are $(n+1)$ mutually exclusive outcomes:

\( \begin{array}{ll}
 s < s_1, & {\rm with\ probability}\ p_0 = 
	0.5\{1 - {\rm erfc}\,(s_1/\sqrt{2})\} \\
 \vdots  & \\
 s \geq s_n, &{\rm with\ probability}\ p_n = 
	0.5 \{{\rm erfc}\,(s_n/\sqrt{2})\}
\end{array} \)

\noindent where 
\begin{equation} 
{\rm erfc}\,(x) \equiv \frac{2}{\sqrt{\pi}} \int_{x}^{\infty} e^{-t^2}
dt.
\end{equation}

Then, if one draws a total of $n$ values from the
distribution, the probability that $k_0$ have values $s < s_1$, $k_1$
have values in the interval $s_1 \leq s < s_2$, \ldots, and that $s_n$
have values $s \geq y_n$, is given by the multinomial distribution:
\begin{equation}
p(k_0,k_1,\ldots,k_n) = \frac{n!}{k_0!\ k_1!\ \cdots\ k_n!}\ p_0^{k_0}
p_1^{k_1} \cdots p_n^{k_n}.
\end{equation}
The probability of a false detection is the sum of all the multinomial
terms which fulfill the original requirement, namely that $k_0 = 0$,
$k_n \geq 1$, $(k_{n-1} + k_n) \geq 2$, and so on: $$P_{false} =
p(0,0,\ldots,n) + p(0,0,\ldots,1,n-1) + \ldots + p(0,1,\ldots,1).  $$

The M33 data cube contains approximately 20500 independent positions,
each of which corresponds to a 220-channel spectrum.  For an
$N_{ch}$-channel spectrum, there are $(N_{ch}-n+1)$ ways of selecting
$n$ adjacent channels.  Thus, over the 94 channel range covering the
velocity interval $-290 <$ \vlsr $ < -80$ there are of order
$N_{trials} = 20500\,(94-n) \sim 2 \times 10^6$ ways of selecting $n$
adjacent channels from the data cube.  If $P_{false} \ll 1$ and
$N_{trials} \gg 1$, the probability of $r$ false detections is given
by Poisson statistics:
\begin{equation}
P(r) = \frac{(NP)^r}{r!} e^{-NP}.
\end{equation}
The probability of any given detection being real is $P_{real} = P(r=0)
= e^{-NP} \sim 1-NP$ for $NP \ll 1$.  If there are a given number of
detections at this significance level, the probability of $q$ of them
being false is simply $P(q)$.

As an example, suppose we detect 3 adjacent channels with $s_1 = s_2 =
s_3 = 2.5\,\sigma$.  In this case the probability of a false detection
is simply the product of the probabilities of a false detection in
each of the 3 channels: $P_{false} = \{0.5\,{\rm
erfc}(2.5/\sqrt{2})\}^3 = 2.39 \times 10^{-7}$.  Using the Poisson
formula, one finds a 60\% likelihood of {\it all} sources at this
significance being real.

If the signal-to-noise ratios of the 3 channels are unequal, then the
probability of a false detection is greater than the simple product of
the 3 probabilities, essentially because one has more than one chance to
draw the less likely values from the population of Gaussian deviates. 
For example, if $s_1 = 2.0\,\sigma,\ s_2 = 2.5\,\sigma$, and $s_3 =
3.4\,\sigma$, 
\begin{equation} P_{false} = p(0,0,0,3) + p(0,0,1,2) + p(0,0,2,1) + 
p(0,1,0,2) + p(0,1,1,1) = 2.39 \times 10^{-7},
\end{equation}
whereas the simple product of the probabilities is $4.76 \times
10^{-8}$.  Again, Poisson statistics calculates about a 60\%
likelihood that all sources at this level are real.  Because of the
non-Gaussian nature of the noise, we ultimately use this probability
as a relative measure for ranking the significance of cloud candidates
(\S\ref{cloudfind}).

\section{GMC--\ion{H}{2} Region Spatial Correlation Function}

To understand in a statistical sense how massive star formation
affects GMCs, it is a matter of interest to determine how many GMCs in
our catalog are physically linked with \ha~emitting \ion{H}{2}
regions. The filling factors $f$ for \ha~ and CO in the survey region
are 2.4\% and 0.41\%, respectively, where both types of emission show
strong positional correspondence with \HI\ filaments.  \ion{H}{2}
regions are so widespread that, to some degree, observed proximity to
GMCs may be random.  If \ion{H}{2} regions and GMCs are all randomly
distributed in a survey area $A_{survey}$, the average separation
$\langle r\rangle$ between a GMC and the nearest \ion{H}{2} region
obeys the relation
\begin{equation}
\frac{\pi\langle r\rangle^{2}f_{\ha}}{\langle
A_{\mathrm{HII}}\rangle}\sim 1,
\end{equation}
where $\langle A_{\mathrm{HII}}\rangle$, the average area of an
optical \ion{H}{2} region, is $\sim 700~\mbox{pc}^{2}$.  If we assume an
\ion{H}{2} region and GMC are closely associated if they are separated
by no more than $r_{\circ}$, say, then the expectation value for the
number of close associations is
\begin{equation}
\label{spat}
N_{exp} = N_{gmc}\left( \frac{r_{\circ}}{\langle r\rangle}\right)^{2}
= \frac{\pi r_{\circ}^{2} A_{survey} f_{\ha} f_{CO}}{\langle A_{CO}\rangle
\langle A_{\mathrm{HII}}\rangle}
\end{equation}
Assuming that $\langle A_{CO}\rangle $ has approximately the same area
as a resolution element of our survey, $\langle
A_{\mathrm{HII}}\rangle /\langle A_{CO}\rangle ~\sim~0.4$, and
$\langle r_{\circ}\rangle ~\sim~ 40$ pc -- the average projected
separation of an \ion{H}{2} region lying at the edge of a GMC -- then
$N_{exp}~\sim 34$.  A comparison of our GMC catalog to the \ion{H}{2}
region catalogs previously referenced yields 100 \ion{H}{2} regions in
close association with GMCs, about a factor of three in excess of what
might be expected if GMC and \ion{H}{2} positions were random.

\end{document}